\documentclass[aps,prc,twocolumn,groupedaddress]{revtex4-1}
\usepackage{graphicx,color}
\def\vc#1{\mbox{\boldmath $#1$}}
\usepackage{longtable}
\usepackage{amsmath}

\begin{document}

\title{Microscopic study of $4\alpha$-particle condensation with proper treatment of resonances}

\author{Y.~\textsc{Funaki}, T.~\textsc{Yamada}$^1$, A.~\textsc{Tohsaki}$^2$, H.~\textsc{Horiuchi}$^{2,3}$, G.~\textsc{R\"opke}$^4$ and P.~\textsc{Schuck}$^{5,6,7}$}


\affiliation{Institute of Physics, University of Tsukuba, Tsukuba 305-8571, Japan}
\affiliation{$^1$Laboratory of Physics, Kanto Gakuin University, Yokohama 236-8501, Japan}
\affiliation{$^2$Research Center for Nuclear Physics (RCNP), Osaka University, Osaka 567-0047, Japan}
\affiliation{$^3$International Institute for Advanced Studies, Kizugawa 619-0225, Japan}
\affiliation{$^4$Institut f\"ur Physik, Universit\"at Rostock, D-18051 Rostock, Germany}
\affiliation{$^5$Institut de Physique Nucl\'eaire, CNRS, UMR 8608, Orsay, F-91406, France}
\affiliation{$^6$Universit\'e Paris-Sud, Orsay, F-91505, France} 
\affiliation{$^7$Laboratoire de Physique et Mod\'elisation des Milieux Condens\'es, CNRS et Universit\'e Joseph Fourier, 25 Av.~des Martyrs, BP 166, F-38042 Grenoble Cedex 9, France}


\date{\today}
\begin{abstract}
The $4\alpha$ condensate state for $^{16}$O is discussed with the THSR (Tohsaki-Horiuchi-Schuck-R\"opke) wave function which has $\alpha$-particle condensate character. Taking into account a proper treatment of resonances, it is found that the $4\alpha$ THSR wave function yields a fourth $0^+$ state in the continuum above the $4\alpha$-breakup threshold in addition to the three $0^+$ states obtained in a previous analysis. It is shown that this fourth $0^+$ $((0_4^+)_{\rm THSR})$ state has an analogous structure to the Hoyle state, since it has a very dilute density and a large component of $\alpha+{^{12}{\rm C}(0_2^+)}$ configuration. Furthermore, single-$\alpha$ motions are extracted from the microscopic 16-nucleon wave function, and the condensate fraction and momentum distribution of $\alpha$ particles are quantitatively discussed. It is found that for the $(0_4^+)_{\rm THSR}$ state a large $\alpha$-particle occupation probability concentrates on a single-$\alpha$ $0S$ orbit and the $\alpha$-particle momentum distribution has a $\delta$-function-like peak at zero momentum, both indicating that the state has a strong $4\alpha$ condensate character. It is argued that the $(0_4^+)_{\rm THSR}$ state is the counterpart of the $0^+_6$ state which was obtained as the $4\alpha$ condensate state in the previous $4\alpha$ OCM (Orthogonality Condition Model) calculation, and therefore is likely to correspond to the $0_6^+$ state observed at $15.1$ MeV.
\end{abstract}

\maketitle

\section{Introduction}\label{sec:intro}

The concept of $\alpha$-clustering is essential to understand the structure of light nuclei~\cite{tang,ikeda68}. In particular, $^{16}$O is one of the nuclei which have the most extensively been studied with the cluster model as well as with the shell model. While the ground state is well understood to have a doubly-closed-shell structure, the excited state, such as the second $0^+$ state, is known as one of  ``the mysterious $0^+$ states'' whose excitation energy, $6.05$ MeV, is too low to be explained in terms of the shell model~\cite{shell}. It had also been mysterious that the second $0^+$ state forms a rotational band with low-lying $2^+$ and $4^+$ states. Besides those states, the third $0^+$ state at $12.05$ MeV has been difficult to be understood from the shell model aspect as well. One of the main aims of the cluster model approach has been to describe these states from a different perspective and, indeed, the studies based on the $\alpha$-cluster model have succeeded in reproducing many experimental data. It was clarified that the $0_2^+$ and $0_3^+$ states have $\alpha(S) + ^{12}$C$(0^+)$ and $\alpha(D)+^{12}$C$(2^+)$ cluster structures, where the $\alpha$ particle rotates in an $S$ and $D$ wave around the $^{12}$C core being in $0^+$ and $2^+$ states, respectively~\cite{Suz76,baye2,Kat92}. The great success  also comes from having further revealed that almost all of the excited states up to $\sim 14$ MeV have considerable large $\alpha + ^{12}$C cluster components~\cite{Suz76,baye2}.

Besides $\alpha + ^{12}$C clustering, $4\alpha$ clustering can also be expected to exist around the decay threshold into $4\alpha$ particles at $14.44$ MeV, as suggested by Ikeda {\it et al.}~\cite{Ike68Hor72}. As an intriguing subject  investigating the $4\alpha$ cluster structure, the presence of a $4\alpha$ linear-chain state still remains an open question~\cite{morinaga,chevallier,freer_chain}. Another possibility are dilute gas-like states composed of weakly interacting $4\alpha$ particles~\cite{thsr}. They are analogues to the Hoyle state, i.e. the $0_2^+$ state in $^{12}$C, well described by a gas-like $3\alpha$-particle structure with dilute density~\cite{carbon}. In good approximation it forms a product state of the $3\alpha$ particles in the lowest $0S$ orbit of the corresponding mean field potential~\cite{thsr,funaki1,YS,matsumura,yamada_12C}. A striking fact is that the Hoyle state wave function obtained more than 30 years ago by solving the $3\alpha$ RGM (Resonating Group Method) equation of motion~\cite{kamimura} is to very good approximation identical to an antisymmetrized product type wave function~\cite{funaki1}, the so-called THSR (Tohsaki, Horiuchi, Schuck and R\"opke) wave function, introduced in Ref.~\cite{thsr}, and slightly modification in Ref.~\cite{funaki_8Be}. Thus, the state has been proposed to be related to boson condensation of $\alpha$ particles in infinite nuclear matter~\cite{slr09,roepke,beyer}. The new interpretation of the Hoyle state as an $\alpha$ condensate has provoked many theoretical and experimental works on $\alpha$-particle condensation phenomena in light nuclei~\cite{itoh,koka,freer,ohkubo,funaki_res,takashina,kurokawa,enyo_12C,neff}. It should be mentioned that also for  $A\neq 4n$ nuclei such as $^{11}$B and $^{13}$C several authors try to find states which are analogous to the Hoyle state~\cite{kawabata,enyo_11B,yamada_11B}.

The first investigation of the $4\alpha$-particle condensate state was performed via the THSR ansatz in Ref.~\cite{thsr}, where two excited $0^+$ states were obtained. In particular, the third $0^+$ ($(0_3^+)_{\rm THSR}$) state has an excitation energy close to the $4\alpha$ threshold and a considerably larger r.m.s. radius, $R_{\rm rms}=3.9$ fm, than the one of the ground state, $R_{\rm rms}=2.7$ fm, and therefore the $(0_3^+)_{\rm THSR}$ state was regarded as a candidate of the $4\alpha$ condensate state. Based on this theoretical prediction, an experimental search for the $4\alpha$ condensate state was undertaken via inelastic $\alpha$ scattering on $^{16}$O. A new $0^+$ state was observed, as the fourth $0^+$ state, at $13.6$ MeV with  a relatively large $\alpha$ decay width of $0.6$ MeV~\cite{wakasa}. The angular distribution of the cross section of the inelastic $\alpha$ scattering to the new state well agrees with the calculated one using the $(0_3^+)_{\rm THSR}$ wave function. The relatively large decay width of $0.6$ MeV is also consistently reproduced by the THSR wave function~\cite{funaki_mpla}. These results lead to the conclusion~\cite{thsr,wakasa,funaki_mpla} that the $(0_3^+)_{\rm THSR}$ state may be assigned to the new $0^+$ state at $13.6$ MeV. Thus the $13.6$ MeV state was considered to be a candidate of the $4\alpha$ condensate state~\cite{wakasa}.

In order to confirm the above mentioned THSR description of $^{16}$O, recently the present authors investigated the $4\alpha$ condensate state via the $4\alpha$ OCM (Orthogonality Condition Model)~\cite{4aocm}. The four-body ($4\alpha$) problem with respect to the $\alpha$-$\alpha$ relative motion was solved in a model space large enough to describe the dilute $4\alpha$ gas-like configuration, as well as $\alpha + ^{12}$C cluster and shell-model-like ground state structures,  using the Gaussian expansion method~\cite{GEM}. A one-to-one correspondence with the observed energy spectrum up to the $0_6^+$ state (see also FIG.~\ref{fig:3}) was observed. It further was shown that the calculated $0_6^+$ ($(0_6^+)_{\rm OCM}$) state has large occupation probability of $61$ \% for $4\alpha$'s in the single $0S$ orbit and also a large overlap with the $\alpha+^{12}$C$^*$(Hoyle state) structure. These results indicate that the $(0_6^+)_{\rm OCM}$ state can be regarded to be the $4\alpha$ condensate state as the analogue to the Hoyle state, to be identified with the sixth experimental $0^+$ state at $15.1$ MeV. The well-known $\alpha$-cluster structures of the $0_2^+$ and $0_3^+$ states were simultaneously reproduced, in agreement with previous OCM calculations~\cite{Suz76, Kat92}. It was also shown that the $0_4^+$ and $0_5^+$ states both have $\alpha+^{12}$C cluster structures, where the $\alpha$ cluster orbits in a higher nodal $S$-wave and in a $P$-wave around the $^{12}$C core with $J^\pi=0_1^+$ and $1_1^-$, respectively. These states had not been discussed so far in the literature.

In this work, we confront our previous analysis with the THSR wave function for $^{16}$O [11] with the above mentioned results from the OCM approach. As we explained above, the $4\alpha$ OCM results tell us that the $0_6^+$ state at $15.1$ MeV is a good candidate for the $4\alpha$ condensate state. In Ref.~\cite{thsr} using the THSR wave function only three $0^+$ states were obtained, including the ground state. This scarcity of $0^+$ states in the THSR approach is easily understood by the fact that it allows only for two limiting configurations, that is a pure Slater determinant for $B=b$ and a pure $\alpha$-particle gas for $B\gg b$~\cite{monopole}. Asymptotic configurations like ${^{12}{\rm C}}(0_1^+)+\alpha$ are absent. They may be represented in a rough average way by the intermediate $0^+$ state. In any case the situation is such that it incited us to reanalyze the previous study of Ref.~\cite{thsr}. The important question is whether or not the counterpart of the $(0_6^+)_{\rm OCM}$ state can also be found with the THSR ansatz, which should be a further state located above the $(0_3^+)_{\rm THSR}$ state. In this respect, we need to recall that, in the previous THSR calculation~\cite{thsr}, a proper resonance boundary condition was not imposed even for the states above the $4\alpha$ threshold. Thus we could not obtain reliable results for higher $0^+$ states than the $(0_3^+)_{\rm THSR}$ state. The other excited state then only mocks up in an average way the states which have dominant $\alpha + ^{12}$C structure.

The purpose of this paper is to report on results of a $4\alpha$ THSR calculation with proper treatment of the resonances. We will see that this calculation gives us a new $0^+$ state as the $(0_4^+)_{\rm THSR}$ state, slightly above the $4\alpha$ threshold. By comparing the wave functions of the $(0_4^+)_{\rm THSR}$ state and the $(0_6^+)_{\rm OCM}$ state in the $4\alpha$ OCM calculation, it will be clarified that both states well correspond to each other. In particular, we will show that the $(0_4^+)_{\rm THSR}$ state has a larger r.m.s. radius, a larger amount of condensate fraction, and a more sharpened peak around zero momentum for the $\alpha$-particle momentum distribution, than the $(0_3^+)_{\rm THSR}$ state. Those properties are comparable to those of the $(0_6^+)_{\rm OCM}$ state. This indicates that the newly found $(0_4^+)_{\rm THSR}$ state should be considered as the $4\alpha$ condensate state and is likely to be assigned to the $0_6^+$ state observed at $15.1$ MeV.

The outline of the present paper is as follows: In Sec. \ref{subsec:thsr}, we introduce the THSR wave function and the microscopic Hamiltonian. The present approach to resonances is briefly explained in Sec.~\ref{subsec:rmsmin}, which is, together with the THSR ansatz, applied to ${^{16}{\rm O}}$ in Sec.~\ref{subsec:04+}. The obtained wave functions of ${^{16}{\rm O}}$ are analyzed in detail in Sects.~\ref{subsec:wf} and ~\ref{subsec:overlap}. The quantities based on $\alpha$-particle bosonic degree of freedom, such as the amount of condensation, are discussed in Sec.~\ref{subsec:boson}. Further discussions are presented in Sects.~\ref{subsec:size} and ~\ref{subsec:interp}. Sec.~\ref{sec:conc} is devoted to the conclusion.

\section{Formulation}\label{sec:form}

\subsection{THSR wave function}\label{subsec:thsr}

In general, the $n\alpha$ cluster model wave function can be presented in the following form:
\begin{equation}
\Phi_{n\alpha}({\vc r}_1, \cdots, {\vc r}_{4n}) = {\cal A}[ \chi({\vc 
R}_1, {\vc R}_2,\cdots, {\vc R}_n) 
\phi_{\alpha_1}\phi_{\alpha_2}\cdots\phi_{\alpha_n}] \label{eq:1}
\end{equation}
with $\cal A$ the antisymmetrizer and $\phi_{\alpha_i}$ the intrinsic wave function of the $i$-th $\alpha$-particle, which is taken as a Gaussian,
\begin{equation}
\phi_{\alpha_i} \propto \exp\Big[-\sum_{1\leq k<l \leq4}({\vc r}_{i,k} - {\vc r}_{i,l})^2/(8b^2)\Big]. \label{eq:2}
\end{equation}
The wave function $\chi$ for the c.o.m. motion of the $\alpha$'s with ${\vc R}_i = \frac{1}{4}({\vc r}_{i,1}+{\vc r}_{i,2}+{\vc r}_{i,3}+{\vc r}_{i,4})$ is, of course, also chosen translationally invariant, that is it depends only on the relative coordinates ${\vc R}_{ij}={\vc R}_i-{\vc R}_j$ or on the corresponding Jacobi coordinates. The spin-isospin part in Eq.~(\ref{eq:1}) is not written out but supposed to be of scalar-isoscalar form. We will not mention it henceforth.

In 2001, Tohsaki, Horiuchi, Schuck, and R\"opke (THSR)~\cite{thsr} adopted, as the $n\alpha$ condensate type wave function, the following ansatz for $\chi$ in Eq.~(\ref{eq:1}),
\begin{eqnarray}
&&\hspace{-1.cm} \chi_{n\alpha}^{\rm THSR}(B;\vc{R}_1,\vc{R}_2,\cdots,\vc{R}_n)\nonumber \\ 
&&\hspace{-0.5cm}=\varphi_0(B;\vc{R}_1-\vc{X}_G) \cdots\varphi_0(B;\vc{R}_n-\vc{X}_G), \label{eq:chi1}
\end{eqnarray}
with the total c.o.m. coordinate $\vc{X}_G=(\vc{R}_1+\cdots+\vc{R}_n)/n$ and $\varphi_0(B;\vc{R}-\vc{X}_G)=\exp(-2(\vc{R}-\vc{X}_G)^2/B^2)$, that is a Gaussian with a large width parameter $B$ which is of the nucleus' dimension. The product of $n$ identical $0S$ wave functions reflects the boson condensate character. This feature is realized as long as the action of the antisymmetrizer in Eq.~(\ref{eq:1}) is sufficiently weak. This type of condensate wave function has been known, in the meantime, to have considerable success, notably with an accurate description of the Hoyle state, proposing it as the first of a series of excited states in $n\alpha$ nuclei with $\alpha$-particle condensate character. Those states can be considered as the precursors of macroscopic $\alpha$-particle condensation in nuclear matter at low density [19].

The internal part of the Hamiltonian we adopt in the present study of $^{16}$O is composed of kinetic energy $-\frac{\hbar^2}{2M}\nabla_i^2$, with nucleon mass $M$, the Coulomb force $V_{ij}^C$ and the effective nuclear interaction: 
\begin{equation}
H=-\sum_{i=1}^{16}\frac{\hbar^2}{2M}\nabla_i^2 - T_G +\sum_{i<j}^{16}V_{ij}^C + \sum_{i<j}^{16} V_{ij}^{(2)} + \sum_{i<j<k}^{16}V_{ijk}^{(3)}, \label{eq:3}
\end{equation}
where the c.o.m. kinetic energy $T_G$ is subtracted. In this calculation, we adopted the same nucleon-nucleon force with two-body $V_{ij}^{(2)}$ and three-body $V_{ijk}^{(3)}$ terms as adopted in Ref.~\cite{thsr}. Both two-body and three-body forces consist of finite-range Gaussian functions with a parameter set named F1 given in Ref.~\cite{f1}.

The wave functions of quantum states in ${^{16}{\rm O}}$ can then  be expanded using the $4\alpha$ THSR wave function, like
\begin{equation}
\Psi_k = \sum_m f_k (B^{(m)}) \Phi_{4\alpha}(B^{(m)}), \label{eq:hwwf}
\end{equation}
where $\Phi_{4\alpha}(B^{(m)})$ is the $4\alpha$ THSR wave function which has the form of
\begin{equation}
\Phi_{4\alpha}(B^{(m)})= {\cal A}[ \chi_{4\alpha}^{\rm THSR}(B^{(m)};\vc{R}_1,\cdots,\vc{R}_4) \phi_{\alpha_1}\cdots\phi_{\alpha_4}]. \label{eq:thsr}
\end{equation}
The discrete variational parameters $B^{(m)}$ represent the generator coordinate of  the Hill-Wheeler ansatz. The expansion coefficients $f_k (B^{(m)})$ and the corresponding eigenenergy $E_k$ for the $k$-th eigenstate are obtained by solving the following Hill-Wheeler equation, 
\begin{eqnarray}
&& \sum_{m^\prime} \left\langle \Phi_{4\alpha}(B^{(m)}) \Big| H-E_k  \Big| \Phi_{4\alpha}(B^{(m^\prime)}) \right\rangle \nonumber \\ 
&& \hspace{4cm}\times f_k (B^{(m^\prime)}) =0.  \label{eq:hw}
\end{eqnarray}

\subsection{Extraction of the resonance wave function}\label{subsec:rmsmin}

One of the aims of the present study is to discuss the states obtained by the THSR ansatz with proper treatment of resonances. The calculation of the resonance state in the bound state approximation is usually done by diagonalizing the Hamiltonian with the use of a finite number of square-integrable basis wave functions. The calculated positive energy eigenstates which are linear combinations of the basis wave functions are divided into resonance states and continuum states.

There are many practical methods to carry out this division. One of the methods, found in Ref.~\cite{rmsmin} by some of the present authors, was shown to give consistent results with the ones obtained with a well known approach to resonances such as the ACCC (Analytic Continuation of the Coupling Constant) method~\cite{accc,TSVL}. We apply the method to the present study, and following the literature~\cite{rmsmin}, briefly explain it in this subsection.

Let us first consider an attractive pseudo-potential $V$ which is added to the original Hamiltonian $H$, yielding
\begin{equation}
H^\prime(\delta) = H + \delta \times V, \label{hamil}
\end{equation}
where $\delta$ is a coupling constant used to vary the strength of the pseudo-potential.  We diagonalize this new Hamiltonian $H^\prime(\delta)$ by using the same set of basis wave functions as used for $H$.  As we increase the coupling constant $\delta$ from the physical value, $\delta=0$, the eigenenergy of any resonance state decreases, and eventually the resonance state is transformed into a bound state. By contrast, continuum states exhibit almost no change in their eigenenergies as $\delta$ increases.
\begin{figure}[htbp]
\begin{center}
\includegraphics[scale=0.75]{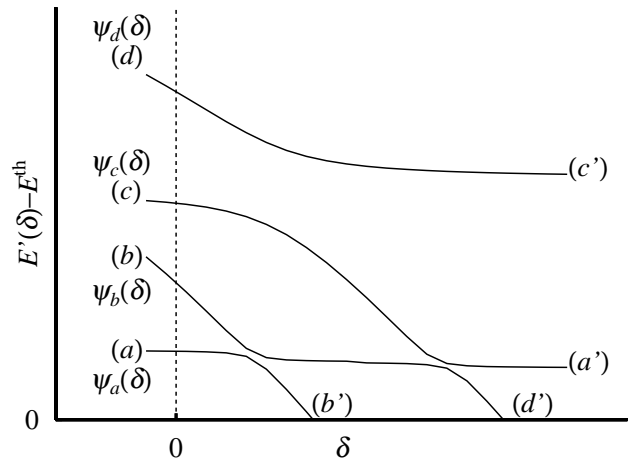}
\caption{Eigenenergies of the Hamiltonian (\ref{hamil}) are schematically shown as functions of $\delta$. There are three crossing points near which the mixing of continuum and resonance components occurs. (See the main text for a detailed explanation.) Figure adopted from Ref.~\cite{rmsmin}.}
\label{fig:scheme}
\end{center}
\end{figure}

 This behavior of the eigenenergies as functions of $\delta$ is shown schematically in FIG.~\ref{fig:scheme}.  In this figure, the curve labeled (b) around $\delta=0$ represents the eigenenergy of the resonance state denoted as $\psi_{\rm b}(\delta)$. It decreases as $\delta$ increases. This curve (b) then crosses the curve (a), which is flat around $\delta=0$ and is the eigenenergy of a continuum state denoted as $\psi_{\rm a}(\delta)$. After the closest approach of the two curves (b) and (a), the curve labeled (b'), which is the smooth continuation of the curve (a), and hence is disconnected from the curve (b), represents the eigenenergy of the resonance state $\psi_{\rm b}(\delta)$. This reminds us of the Landau-Zener level crossing scenario, see e.g.~\cite{Ring/Schuck}. The eigenenergy of the continuum state $\psi_{\rm a}(\delta)$, which is first represented by the curve (a), continues with an almost flat shape, except near the crossing points. Eventually it is converted into the curve (a'). In the case of the resonance state $\psi_{\rm b}(\delta)$ at $\delta=0$, we can consider that this state contains a negligible amount of admixture of continuum state components, since it is located far from the crossing points with the continuum states.

However, in the case of the resonance state $\psi_{\rm d}(\delta)$ corresponding to the resonance state curve (d) in FIG.~\ref{fig:scheme}, the state is at $\delta=0$ already involved in the crossing with the continuum state curve (c). Therefore, $\psi_{\rm d}(\delta)$ at $\delta=0$ should contain a sizable admixture of the continuum state. This is the usual situation for a resonance state with a broad width, and the resonance wave function obtained by using the bound state approximation cannot be trusted, because of the sizable admixture of the continuum state.

In order to overcome this difficulty, we note the utility of the continuum state wave functions obtained using the bound state approximation in the neighborhood of the resonance state, since they should contain sizable admixtures of the resonance state component. In our present simple example considered in FIG.~\ref{fig:scheme}, we notice the utility of the continuum state wave function $\psi_{\rm c}(\delta)$. We can express both states,  $\psi_{\rm d}(\delta)$ and $\psi_{\rm c}(\delta)$, as linear combinations of the pure resonance state $\Psi_{\rm d}(\delta)$ and the pure continuum state $\Psi_{\rm c}(\delta)$: 
\begin{equation}
\psi_{\rm d}(\delta) = \alpha \Psi_{\rm d}(\delta) + 
\beta \Psi_{\rm c}(\delta), \quad  
\psi_{\rm c}(\delta) = -\beta \Psi_{\rm d}(\delta) + 
\alpha \Psi_{\rm c}(\delta). 
\end{equation}
The method for the extraction of the pure resonance component $\Psi_{\rm d}(\delta)$ which we proposed in Ref.~\cite{rmsmin} is obtained by noticing that the extension of the density of a resonance state is far smaller than the one of continuum states. More precisely, we can determine $\alpha$ and $\beta$ from the requirement that the quantity
\begin{equation}
\langle \Psi_{\rm d}(\delta) | \sum_{j=1}^A (\vc{r}_j-\vc{X}_G)^2 | \Psi_{\rm d}(\delta) \rangle
\end{equation}
be a minimum, where
\begin{equation}
\Psi_{\rm d}(\delta) = \alpha \psi_{\rm d}(\delta) - \beta 
\psi_{\rm c}(\delta).
\end{equation} 
Here $A$ is the mass number. This requirement is satisfied by diagonalizing the operator $\sum_{j=1}^A (\vc{r}_j-\vc{X}_G)^2$ in the basis states $\psi_{\rm d}(\delta)$ and $\psi_{\rm c}(\delta)$, and by choosing $(\alpha, -\beta)$ to be the eigenvector belonging to the smallest eigenvalue.

When the resonance component is distributed among several energy eigenstates, we need to diagonalize the operator $\sum_{j=1}^A (\vc{r}_j-\vc{X}_G)^2$ in these energy eigenstates. When the energy eigenstates contain two resonance components, the diagonalization of the operator $\sum_{j=1}^A (\vc{r}_j-\vc{X}_G)^2$ gives two small eigenvalues.  Two resonance states are obtained by diagonalizing the Hamiltonian $H^\prime(\delta)$ with the two wave functions which give the two smallest eigenvalues of the operator $\sum_{j=1}^A (\vc{r}_j-\vc{X}_G)^2$.

We are now able to calculate approximate values of resonance energies along the resonance energy curve. If a point on the resonance state curve is not influenced by any crossing points with continuum state curves, we can regard the energy of the point to be just the resonance energy.  On the other hand, if a point on the resonance state curve is influenced by some near-by crossing point, we extract the resonance wave function using the procedure explained above and then calculate the expectation value of the Hamiltonian $H^\prime(\delta)$ with the extracted wave function, which is just the resonance energy at this point. We apply this method for extracting the $(0_4^+)_{\rm THSR}$ state with the use of the THSR wave function in Sec.~\ref{subsec:04+}.

\section{Results and discussions}\label{sec:wf}

\subsection{Extraction of the resonance components for the $(0_4^+)_{\rm THSR}$ state}\label{subsec:04+}

In the Hamiltonian~(\ref{eq:3}), the explicit form of the long-range part of two-body nuclear force $V_{ij}^{(2)}$ is given by
\begin{equation}
V=-5[{\rm MeV}]\exp \Big[-\Big(\frac{r_{ij}}{2.5[{\rm fm}]}\Big)^2\Big]\Big(1-M_1^{(2)}+M_1^{(2)}P_{ij}\Big),
\end{equation}
with the value of Majorana parameter $M_1^{(2)}=0.75$ and $P_{ij}$ the exchange operator of spatial coordinates. Suppose that we here artificially vary the parameter $M_1^{(2)}$ around the physical value $0.75$, or equivalently, vary the coupling constant defined as $\delta \equiv -M_1^{(2)}+0.75$ around $\delta=0$. This corresponds to the situation explained in Sec.~\ref{subsec:rmsmin}, where a pseudo potential $V_{ij}^{(PS)}$ is added to the original Hamiltonian, like 
\begin{equation}
H^\prime(\delta)=H+ \delta \sum_{i<j}V_{ij}^{(PS)}, \label{eq:32}
\end{equation}
with
\begin{equation}
V_{ij}^{(PS)}= -5[{\rm MeV}] \exp \Big[-\Big(\frac{r_{ij}}{2.5[{\rm fm}]}\Big)^2\Big](1-P_{ij}).
\end{equation}

The energy eigenvalues and eigenfunctions can then be obtained as $\delta$-dependent solutions of the following Hill-Wheeler equation,
\begin{eqnarray}
&& \sum_{m^\prime} \left\langle \Phi_{4\alpha}(B^{(m)}) \Big| H(\delta)-E_k(\delta) \Big| \Phi_{4\alpha}(B^{(m^\prime)}) \right\rangle \nonumber \\
&& \hspace{4cm}\times f_k (B^{(m^\prime)},\delta ) =0.  \label{eq:4}
\end{eqnarray}
Here the $k$-th eigenfunction is given as,
\begin{equation}
\Psi_k(\delta) = \sum_m f_k (B^{(m)},\delta) \Phi_{4\alpha}(B^{(m)}). \label{eq:5}
\end{equation}

In this calculation, for any values of $\delta$, the following $20$ points of the generator coordinate $B^{(m)}=\sqrt{b^2+2(R_0^{(m)})^2}$ are adopted: $R_0^{(m)}=0.5\ r_0^{m-1}$ with $r_0= 1.196$ and $m=1,\cdots,20$. The size parameter of the $\alpha$ particle is fixed as $b=1.44$ fm.

\begin{figure}[htbp]
\begin{center}
\includegraphics[scale=0.77]{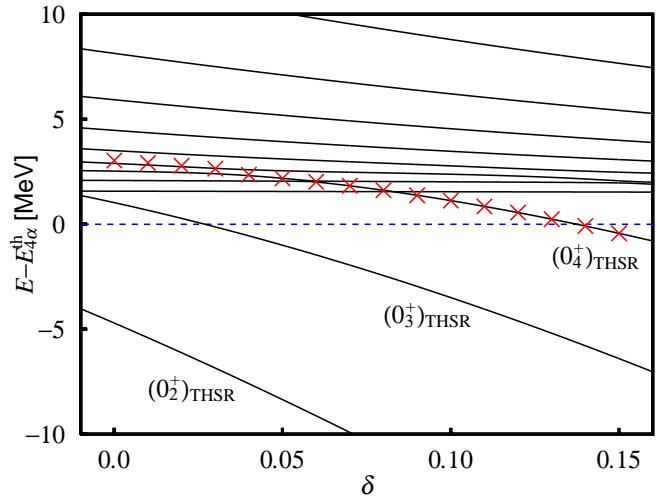}
\caption{(color online). The energy eigenvalues of the Hill-Wheeler Eq.~(\ref{eq:4}) (solid curves) which are measured from the $4\alpha$ threshold (doted curve). The crosses denote the binding energies of the fourth $0^+$ state $\phi_R(\delta)$ in Eq.~(\ref{eq:spps}).}\label{fig:2}
\end{center}
\end{figure}
\begin{figure}[htbp]
\begin{center}
\includegraphics[scale=0.77]{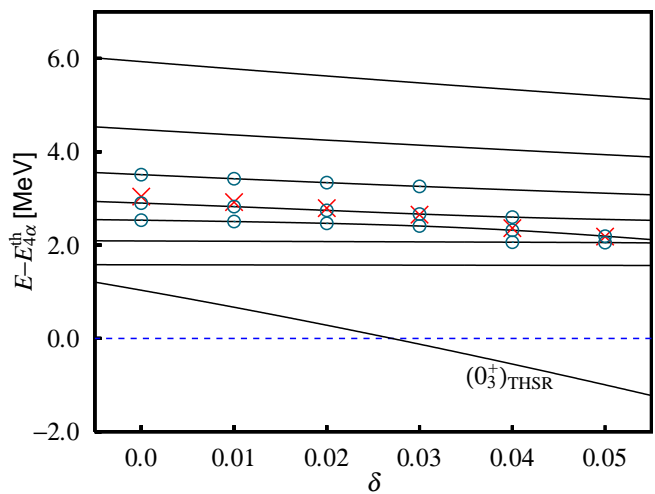}
\caption{(color online). The region of $0.0 \leq \delta \leq 0.05$ of FIG.~\ref{fig:2} is enlarged. The adopted eigenstates in the superposition of Eq.~(\ref{eq:spps}) are denoted by open circles. The crosses denote the binding energies of the fourth $0^+$ state $\phi_R(\delta)$ in Eq.~(\ref{eq:spps}).}\label{fig:enlarge}
\end{center}
\end{figure}

In FIG.~\ref{fig:2}, the binding energies $E_k(\delta)-E_{4\alpha}^{\rm th}$, with $E_{4\alpha}^{\rm th}$ the $4\alpha$ threshold energy, are plotted up to 10 MeV in the range $-0.01 \leq \delta \leq 0.16$. The $4\alpha$ threshold energy is calculated as $E_{4\alpha}^{\rm th}=4E_\alpha = -112$ MeV, with $E_\alpha$ the binding energy of the $\alpha$ particle, which does not depend on values of $\delta$, since the pseudo potential $V_{ij}^{(PS)}$ does not contribute to the binding energy of the $\alpha$ particle with the intrinsic wave function $\phi_\alpha$ in Eq.~(\ref{eq:2}). With the increase of $\delta$ from left to right, we see that three curves go down and pass through the $4\alpha$ threshold one after another. The curves are considered to be trajectories corresponding to a resonance or bound state, which has a compact structure and is sensitive to the  variation of the strength of the pseudo potential, as discussed in Sec.~\ref{subsec:rmsmin}. On the contrary, the other curves, which are relatively straight, i.e. insensitive to the variation, are considered to correspond to continuum states with unrestrictedly large spatial extension.

We should here notice the important fact that the fourth $0^+$ state should be at around $3$ MeV above threshold and that it has been missed in the previous calculation of Ref.~\cite{thsr}. With the decrease of $\delta$, the trajectory for the fourth $0^+$ state passes through the $4\alpha$ threshold at $\delta \simeq 0.14$, and then from $\delta \simeq 0.03$ the trajectory quickly becomes invisible. This is because the fourth state has non-negligible decay width as a resonance. 


This is the very reason why the fourth $0^+$ state could not be found in Ref.~\cite{thsr}, while the first, second, and third $0^+$ states can without problem be assigned to the single eigenstates of the Hamiltonian, $\Psi_1(\delta=0)\equiv \Psi_1$, $\Psi_2(\delta=0)\equiv \Psi_2$ and $\Psi_3(\delta=0)\equiv \Psi_3$, in Eq.~(\ref{eq:5}), respectively.

However, the method explained in Sec.~\ref{subsec:rmsmin} enables us to extract the fourth state, which must exist in our approach around $3$ MeV above the $4\alpha$ threshold. The fourth $0^+$ wave function can then be expressed as
\begin{equation}
\phi_R(\delta)= \sum_k D_k \Psi_k(\delta), \label{eq:spps}
\end{equation}
where the coefficients $D_k$ are determined so that the correct resonance wave function $\phi_R(\delta)$ should have a smaller radius than the continuum wave functions with spatially extended structures: $\phi_R(\delta)$ should satisfy the following relation against the variation of $\{D_k \}$,
\begin{equation}
\frac{d}{d D_k} \langle \phi_R(\delta)|\sum_{i=1}^{16} (\vc{r}_i-\vc{X}_G)^2 | \phi_R(\delta) \rangle =0.
\end{equation}

In FIG.~\ref{fig:enlarge}, where the region $0.0 \leq \delta \leq 0.05$ in FIG.~\ref{fig:2} is enlarged, the adopted eigenstates are marked by open circles at given $\delta$ values, and the binding energy of the extracted resonance state is denoted by crosses. For example, at $\delta=0$, we utilize in the superposition of Eq.~(\ref{eq:spps}) the sixth, seventh and eighth states: $\Psi_6(\delta=0)$, $\Psi_7(\delta=0)$ and $\Psi_8(\delta=0)$ obtained by solving the Hill-Wheeler equation~(\ref{eq:4}), the binding energies of which are marked by three open circles, while the cross at $\delta=0$ denotes the binding energy of the extracted fourth $0^+$ state. We can also see in this figure that binding energies of the extracted resonance states denoted by crosses are reasonably on the trajectory for the fourth $0^+$ state in a wide range of $\delta$ values, indicating the validity of this treatment of resonances.

In all subsequent sections, as a matter of convenience, we refer to the wave function of the extracted fourth $0^+$ state at $\delta=0$ as simply $\Psi_4$ instead of $\phi_R(\delta=0)$. Thus, in the following, we represent the wave functions of $(0_1^+)_{\rm THSR}$, $(0_2^+)_{\rm THSR}$, $(0_3^+)_{\rm THSR}$, and $(0_4^+)_{\rm THSR}$ states as $\Psi_1$, $\Psi_2$, $\Psi_3$, and $\Psi_4$, respectively.

\subsection{Features of the $(0_1^+)_{\rm THSR}$--$(0_4^+)_{\rm THSR}$ states}\label{subsec:wf}

\begin{figure}[htbp]
\begin{center}
\includegraphics[scale=1.]{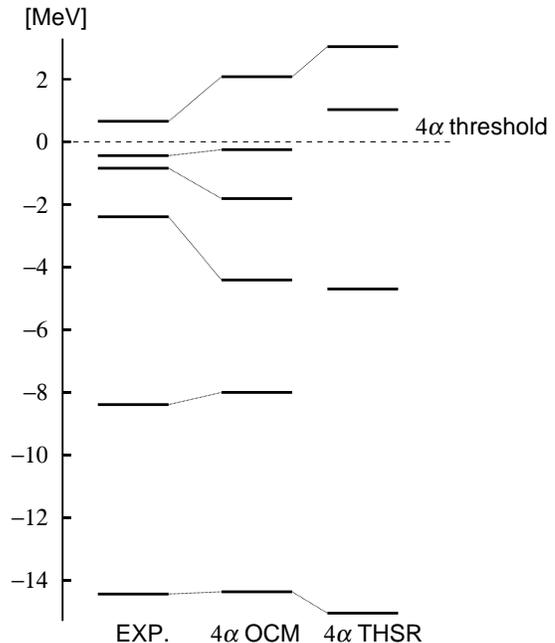}
\caption{Comparison of the $0^+$ energy spectra between experiment, the $4\alpha$ OCM calculation~\cite{4aocm} and the present calculation. Dotted line denotes the $4\alpha$ threshold. Experimental data are taken from Ref.~\cite{ajze} and from Ref.~\cite{wakasa} for the $0_4^+$ state. See Table~\ref{tab:1} for the values.}\label{fig:3}
\end{center}
\end{figure}

We present the calculated spectrum for the $(0_1^+)_{\rm THSR}$--$(0_4^+)_{\rm THSR}$ states in FIG.~\ref{fig:3}, together with the results of experiment and the  $4\alpha$ OCM calculation~\cite{4aocm}. We can see that the $4\alpha$ OCM calculation gives a satisfactory one-to-one correspondence with the experimental spectrum. This can be understood by the fact that in the calculation the relative motions of $\alpha$ particles are solved in a huge model space, which is spanned by Gaussian basis functions~\cite{GEM}. This allows us to represent the $4\alpha$-particle gas, $\alpha+^{12}$C clustering, and shell-model configurations, which are realized in the $(0_6^+)_{\rm OCM}$ state, the $(0_2^+)_{\rm OCM}$--$(0_5^+)_{\rm OCM}$ states, and the ground state $(0_1^+)_{\rm OCM}$, respectively. Thus, the $(0_6^+)_{\rm OCM}$ state, which has the $\alpha$ condensate character, is to be identified with the experimental $0_6^+$ state at $15.1$ MeV as the $4\alpha$ condensate state. We should note that the observed widths of the experimental $0_4^+$, $0_5^+$, and $0_6^+$ states also consistently correspond to the calculated ones (see Table~\ref{tab:1}). In the same table, we can see that the $M(E0)$ values for the $(0_2^+)_{\rm OCM}$, $(0_3^+)_{\rm OCM}$, and $(0_5^+)_{\rm OCM}$ states are in good agreement with the corresponding experimental values.

On the other hand, the present THSR ansatz gives only four $0^+$ states including the ground state. Let us now recall that in principle the THSR wave function of Eq.~(\ref{eq:thsr}) allows for only two limiting configurations, a pure Slater determinant for $B=b$ and a pure $\alpha$-particle gas for $B\gg b$. Thus, as already mentioned, the scarcity of the $0^+$ states comes from the fact that the THSR wave functions can represent the asymptotic configurations like the $\alpha+^{12}$C clustering only in a rough average way. In particular, the $0_3^+$ and $0_5^+$ states which are considered to mainly include $\alpha(D)+^{12}$C$(2_1^+)$ and $\alpha(P)+^{12}$C$(1_1^-)$ clustering configurations, respectively, can hardly be represented by this wave function, since it does not possess nonzero spin waves with respect to the $\alpha$-$\alpha$ relative motions. We should, however, note that the THSR wave functions are appropriate for describing states which dominantly have a gas-like structure of $\alpha$ particles weakly interacting in a relative $S$-wave, as the $\alpha$ condensate. Also the ground state with essentially a Slater determinant configuration is within the possibilities of the THSR wave function.

\begin{table*}
\caption{The binding energies $E-E^{\rm th}_{4\alpha}$, r.m.s. radii $R_{\rm rms}$, monopole matrix elements $M(E0)$, and $\alpha$ decay widths $\Gamma$, in units of MeV, fm, fm$^2$, and MeV, respectively, where $E^{\rm th}_{4\alpha}=4E_{\alpha}$ denotes the $4\alpha$ threshold energy, with $E_\alpha$ the binding energy of the $\alpha$ particle. The values of the previous $4\alpha$ OCM calculation~\cite{4aocm} and of experiment~\cite{ajze,wakasa} are also shown.}\label{tab:1}\begin{tabular}{cccccllccccclllcccc}
\hline\hline
 &  & \multicolumn{2}{c}{THSR} &  &  &  &  &  & \multicolumn{2}{c}{$4\alpha$ OCM} &  &  &  &  &  & \multicolumn{2}{c}{Experiment} &  \\
 & $E-E^{\rm th}_{4\alpha}$ & $R_{\rm rms}$ & $M(E0)$ & $\Gamma$ &  &  &  & $E-E^{\rm th}_{4\alpha}$ & $R_{\rm rms}$ & $M(E0)$ & $\Gamma$ &  &  &  & $E-E^{\rm th}_{4\alpha}$ & $R_{\rm rms}$ & $M(E0)$ & $\Gamma$ \\
\hline
$(0_1^+)_{\rm THSR}$ & $-15.05$ & 2.5 &  &  &  &  & $(0_1^+)_{\rm OCM}$ & $-14.37$ & 2.7 &  &  &  &  &  & $-14.44$ & $2.71 \pm 0.02$ &  &  \\
$(0_2^+)_{\rm THSR}$ & $-4.7$ & 3.1 & 9.8 &  &  &  & $(0_2^+)_{\rm OCM}$ & $-8.00$ & 3.0 & 3.9 &  &  &  &  & $-8.39$ &  & $3.55 \pm 0.21$ &  \\
 &  &  &  &  &  &  & $(0_3^+)_{\rm OCM}$ & $-4.41$ & 3.1 & 2.4 &  &  &  &  & $-2.39$ &  & $4.03 \pm 0.09$ &  \\
$(0_3^+)_{\rm THSR}$ & $1.03$ & 4.2 & 2.5 & 1.6 &  &  & $(0_4^+)_{\rm OCM}$ & $-1.81$ & 4.0 & 2.4 & $\sim 0.15$ &  &  &  & $-0.84$ &  & no data & 0.6 \\
 &  &  &  &  &  &  & $(0_5^+)_{\rm OCM}$ & $-0.248$ & 3.1 & 2.6 & $\sim 0.05$ &  &  &  & $-0.43$ &  & $3.3 \pm 0.7$ & 0.185 \\
$(0_4^+)_{\rm THSR}$ & $3.04$ & 6.1 & 1.2 & 0.14 &  &  & $(0_6^+)_{\rm OCM}$ & $2.08$ & 5.6 & 1.0 & $\sim 0.05$ &  &  &  & $0.66$ &  & no data & 0.166 \\
\hline\hline
 &  &  &  &  &  &  &  &  &  &  &  &  &  &  &  &  &  &  \\
\end{tabular}
\end{table*}

As mentioned above, we concluded in Ref.~\cite{4aocm} that the $(0_6^+)_{\rm OCM}$ state, which might correspond to the experimental $0_6^+$ state at $15.1$ MeV, may be identified as the $4\alpha$ condensate state. It is therefore important to see whether or not the THSR ansatz can give a state corresponding to the $(0_6^+)_{\rm OCM}$ state and the $0_6^+$ state at $15.1$ MeV state as well. On the other hand, as mentioned in Sec.~\ref{sec:intro}, it was discussed in Ref.~\cite{wakasa} that the $(0_3^+)_{\rm THSR}$ state could be assigned to the $0_4^+$ state observed at 13.6 MeV, which corresponds to the $(0_4^+)_{\rm OCM}$ state. This means that we should now examine the $(0_4^+)_{\rm THSR}$ state found in the preceding subsection, as an important candidate for the $4\alpha$ condensate state.

In Table~\ref{tab:1}, the binding energies, r.m.s. radii, monopole matrix elements with the ground state $M(E0)$, and $\alpha$-decay widths of the $(0_1^+)_{\rm THSR}$--$(0_4^+)_{\rm THSR}$ states are shown. The reason why the present numbers for the $(0_1^+)_{\rm THSR}$--$(0_3^+)_{\rm THSR}$ states differ from those in Ref.~\cite{thsr} is due to the difference of the treatment of c.o.m. motion in the THSR wave function. In the present work the c.o.m. motion is completely eliminated, while in Ref~\cite{thsr} this is approximately done. The $(0_3^+)_{\rm THSR}$ state has a large r.m.s. radius of $4.2$ fm and the $(0_4^+)_{\rm THSR}$ state has even a larger one of $6.1$ fm, which are comparable to the values for the $(0_4^+)_{\rm OCM}$ and $(0_6^+)_{\rm OCM}$ states in the $4\alpha$ OCM calculation, respectively.

The $(0_4^+)_{\rm THSR}$ state as well as the $(0_6^+)_{\rm OCM}$ state come quite a bit higher in energy than the observed $0_6^+$ state at 15.1 MeV. This probably entails that these states have the extremely large r.m.s. radii, $6.1$ fm and $5.6$ fm, respectively. These two features possibly go together, since a state high up in the Coulomb barrier will necessarily have a quite more spatially extended structure. 

The $M(E0)$ values of the $(0_3^+)_{\rm THSR}$ and $(0_4^+)_{\rm THSR}$ states well agree with those of the $(0_4^+)_{\rm OCM}$ and $(0_6^+)_{\rm OCM}$ states, respectively. These indicate that the $(0_4^+)_{\rm THSR}$ state corresponds to the $(0_6^+)_{\rm OCM}$ state, and hence to the $15.1$ MeV state. It will be discussed in Sec.~\ref{subsec:interp}, to what extent the $(0_3^+)_{\rm THSR}$ state approximates the $(0_4^+)_{\rm OCM}$ state.


\begin{figure}[htbp]
\begin{center}
\includegraphics[scale=0.75]{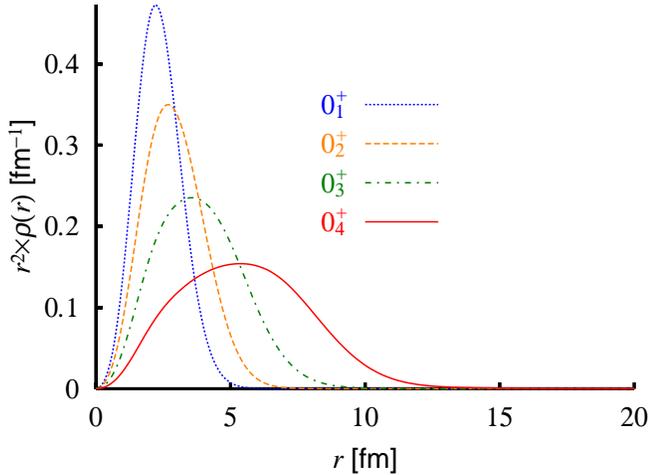}
\caption{(color online). Nucleon density distributions of the $(0_1^+)_{\rm THSR}$--$(0_4^+)_{\rm THSR}$ states defined by Eq.~(\ref{eq:density3}). Note that they are multiplied by $r^2$.}\label{fig:4}
\end{center}
\end{figure}
\begin{figure*}[htbp]
\begin{center}
\includegraphics[scale=0.7]{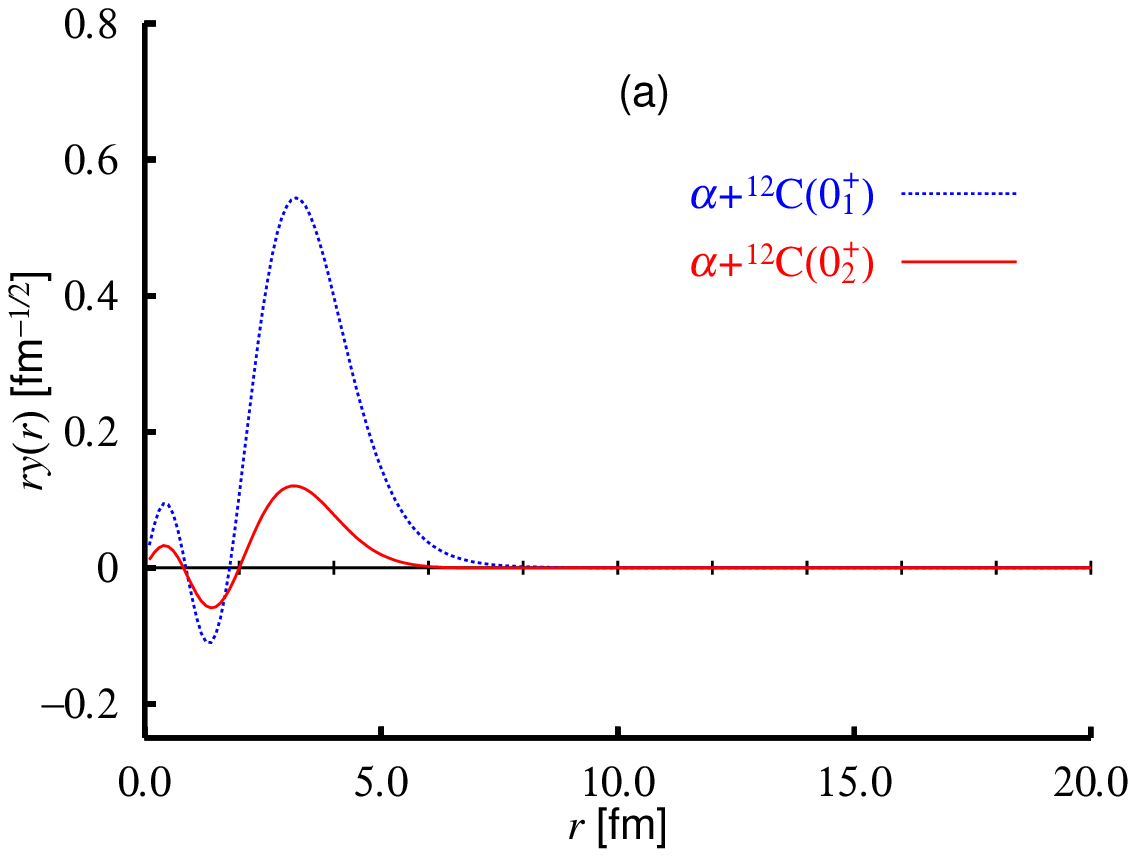}
\includegraphics[scale=0.7]{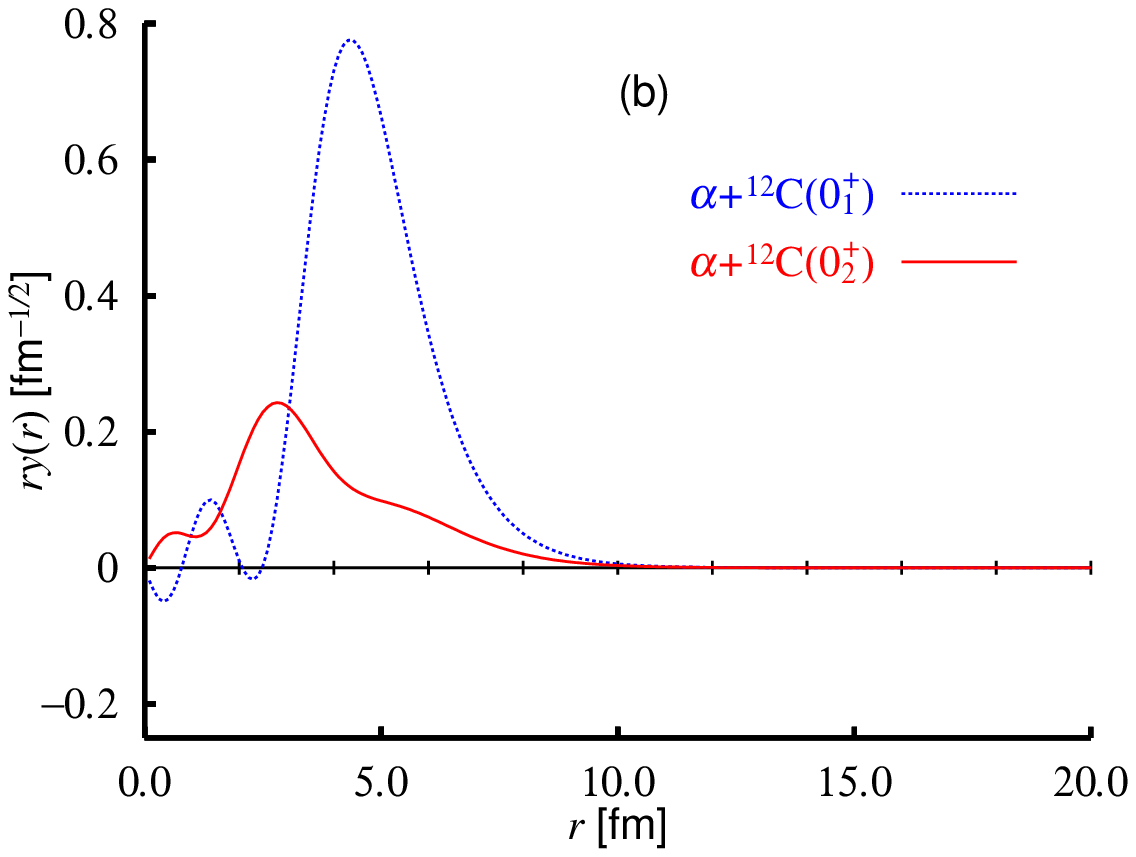}
\end{center}
\begin{center}
\includegraphics[scale=0.7]{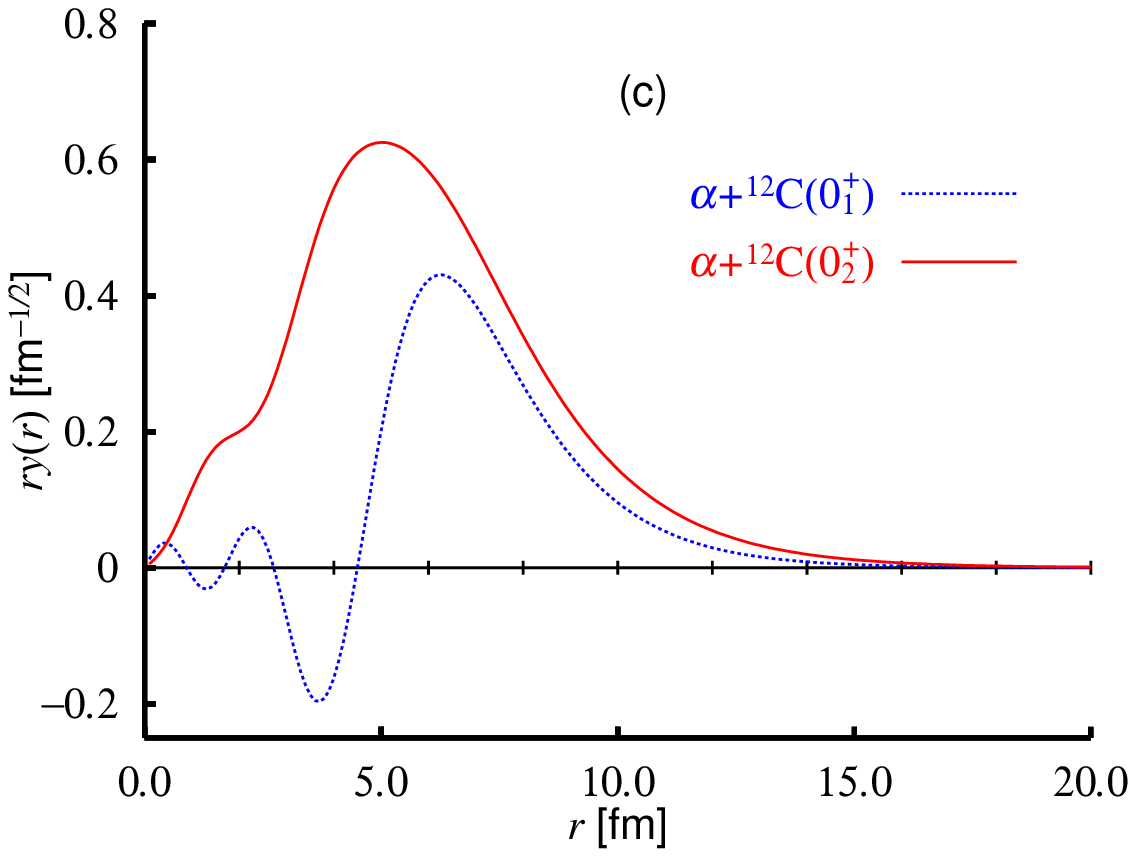}
\includegraphics[scale=0.7]{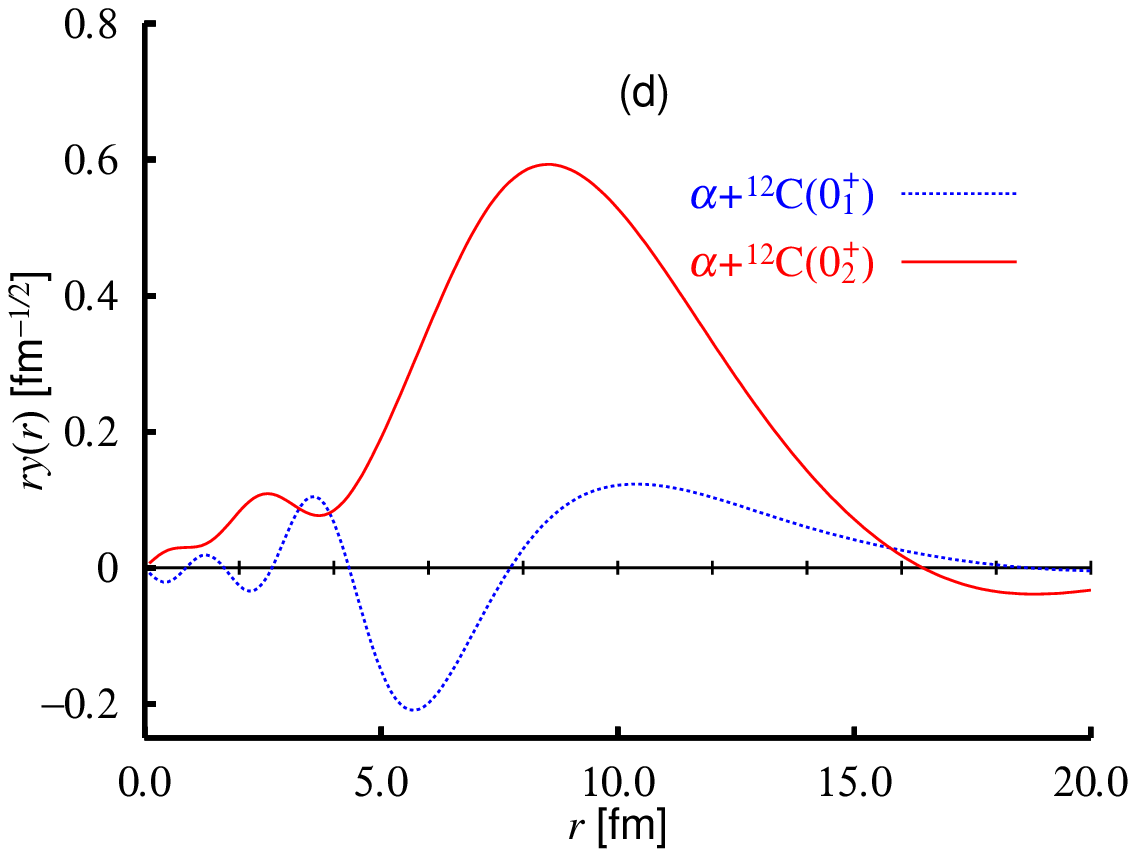}
\end{center}
\caption{(color online). RWA's $r{\cal Y}_{L=0}(r)$ defined by Eq.~(\ref{eq:rwa}) for (a) the $(0_1^+)_{\rm THSR}$, (b) $(0_2^+)_{\rm THSR}$, (c) $(0_3^+)_{\rm THSR}$, and (d) $(0_4^+)_{\rm THSR}$ states in two channels $\alpha + {^{12}{\rm C}}(0_1^+)$ (dotted curve) and $\alpha + {^{12}{\rm C}}(0_2^+)$ (solid curve).}\label{fig:5}
\end{figure*}

The nucleon density distribution defined by,
\begin{equation}
\varrho(r)=\big\langle \Psi_k \big| \frac{1}{16}\sum_{i=1}^{16}\delta(|\vc{r}_i-\vc{X}_G|-r) \big| \Psi_k \big\rangle, \label{eq:density3}
\end{equation}
for the $(0_1^+)_{\rm THSR}$--$(0_4^+)_{\rm THSR}$ states are shown in FIG.~\ref{fig:4}. While the ground state has a compact saturation density, the  $(0_4^+)_{\rm THSR }$ state has a very dilute density structure. This is a common feature of Hoyle-like states. Since the effect of the antisymmetrizer in the THSR wave function will be negligible in the state with dilute density~\cite{funaki_antisymmet}, this state is considered to have the product structure of loosely bound $4\alpha$ particles, contained in the THSR wave function.

\begin{figure}[htbp]
\begin{center}
\includegraphics[scale=0.7]{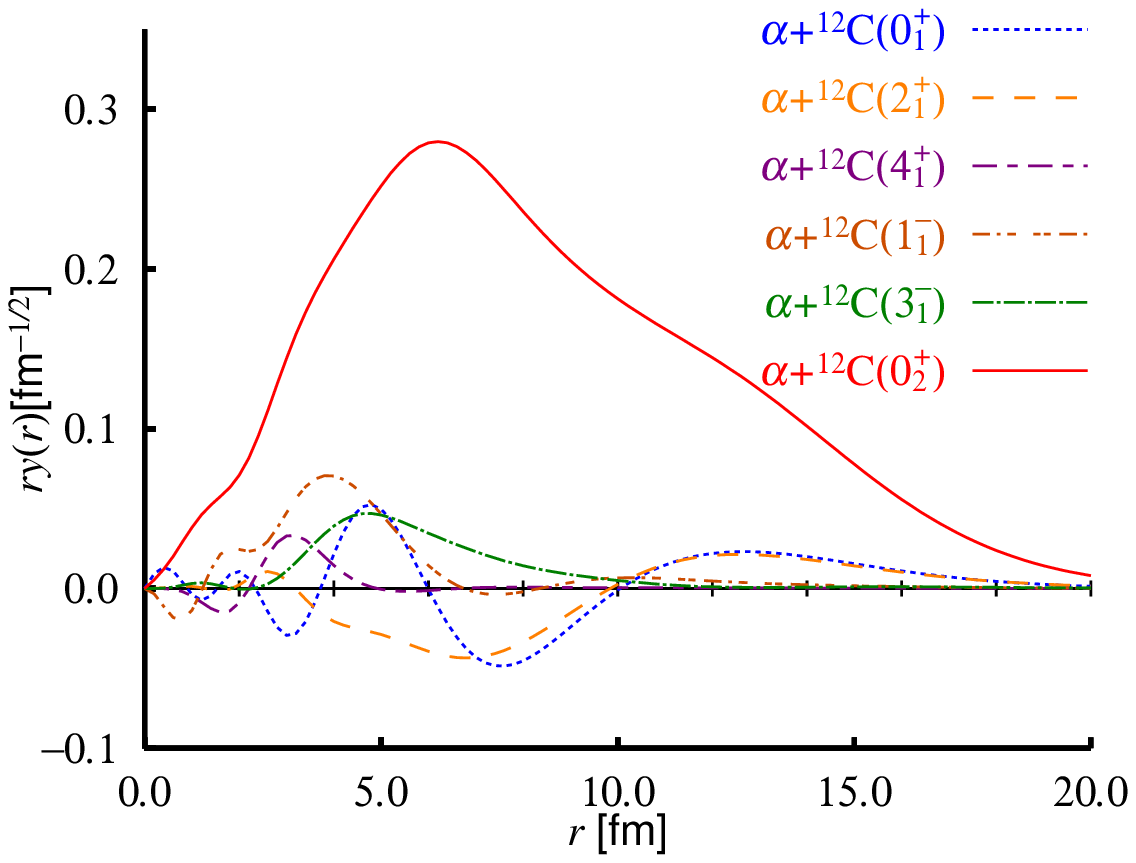}
\caption{(color online). RWA for the $(0_6^+)_{\rm OCM}$ calculated in Ref.~\cite{4aocm}.}\label{fig:rwa_ocm}
\end{center}
\end{figure}

The analogy to the Hoyle state can be discussed directly in the $^{16}$O system by calculating the overlap amplitude with the $\alpha + ^{12}$C$(0_2^+)$ structure. The $4\alpha$ condensate state must have a large overlap with this structure, since it is natural that if one $\alpha$ particle is knocked out from the $4\alpha$ condensate, the remaining $^{12}$C nucleus is in the Hoyle state with the $3\alpha$ condensate structure. This can be seen in FIGS.~\ref{fig:5}(a)--(d), where the following reduced width amplitudes (RWA's) for the $(0_1^+)_{\rm THSR }$--$(0_4^+)_{\rm THSR }$ states are calculated,
\begin{equation}
{\cal Y}_{L=0}(r)=\Big\langle \Big[ \frac{\delta(r^\prime - r)}{r^{\prime 2}}Y_{00}(\vc{\hat r}^\prime)\Psi({^{12}{\rm C}}) \Big] \Big|\Psi_k \Big\rangle , \label{eq:rwa}
\end{equation}
with $\Psi({^{12}{\rm C}})$ being in the ground or Hoyle state of $^{12}$C obtained via THSR ansatz. For the ground state $(0_1^+)_{\rm THSR}$, the RWA in the channel $\alpha + ^{12}$C$(0_1^+)$ has a two-nodal behavior due to the Pauli principle. Accordingly, the $(0_2^+)_{\rm THSR}$, $(0_3^+)_{\rm THSR}$ and $(0_4^+)_{\rm THSR}$ states have RWA's with three, four and five nodes, respectively. The Hoyle state component is then the most largely included in the $(0_4^+)_{\rm THSR}$ state (see FIG.~\ref{fig:5}(d)), though the $(0_3^+)_{\rm THSR}$ state also has a relatively large Hoyle-state component (see FIG.~\ref{fig:5}(c)). We show in FIG.~\ref{fig:rwa_ocm} the RWA's for the $(0_6^+)_{\rm OCM}$ state. We can see that the amplitudes in the ground state $\alpha+{^{12}{\rm C}}(0_1^+)$ and the Hoyle state $\alpha+{^{12}{\rm C}}(0_2^+)$ channels, for the $(0_6^+)_{\rm OCM}$ and $(0_4^+)_{\rm THSR}$ states, well correspond to each other. This indicates that the $(0_4^+)_{\rm THSR}$ state is the counterpart of the $(0_6^+)_{\rm OCM}$ state and 
most appropriate for the $4\alpha$ condensate state, rather than the $(0_3^+)_{\rm THSR}$ state. On the contrary, the $(0_1^+)_{\rm THSR}$ and $(0_2^+)_{\rm THSR}$ states dominantly have a large amplitude in the ground state channel $\alpha + ^{12}$C$(0_1^+)$, reflecting rather compact structures for these states. 

Based on the $R$-matrix theory~\cite{lane}, we can calculate the $\alpha$-decay widths $\Gamma_L$ for the $(0_3^+)_{\rm THSR}$ and $(0_4^+)_{\rm THSR}$ states, which are shown in Table~\ref{tab:1}, with the following fomulae: 
\begin{eqnarray}
\begin{array}{c}
\Gamma_L =2P_L(a) \gamma^2_L(a), \vspace{0.6em} \\
P_L(a)=\dfrac{ka}{F_L^2(ka)+G_L^2(ka)}, \vspace{0.6em}  \\
\gamma^2_L(a)=\theta^2_L(a)\gamma^2_{\rm W}(a), \vspace{0.6em} \\ 
\gamma^2_{\rm W}(a)=\dfrac{3\hbar^2}{2\mu a^2},  \\ 
\end{array}
\label{eq:gamma}
\end{eqnarray}
where $k$, $a$ and $\mu$ are the wave number of the relative motion, the channel radius, and the reduced mass, respectively, and $F_L$, $G_L$, and $P_L(a)$ are the regular and irregular Coulomb wave functions and the corresponding penetration factor, respectively. The reduced width of $\theta^2_L(a)$ is related to the RWA's as, $\theta^2_L(a)=\frac{a^3}{3}{\cal Y}_L^2(a)$. In this calculation, the decay energies are all taken as given by the experimental values: The excitation energy of the $(0_4^+)_{\rm THSR}$ state is assumed to be the one of the observed $0_6^+$ state at $15.1$ MeV. Similarly the excitation energy of the $(0_3^+)_{\rm THSR}$ state is assumed to be the one of the observed $0_4^+$ state at $13.6$ MeV. The total $\alpha$-decay width of $0.14$ MeV for the $(0_4^+)_{\rm THSR}$ state is in good agreement with the corresponding experimental value of $0.17$ MeV. This total width only comes from the decay into the $\alpha + {^{12}{\rm C}}(0_1^+)$ channel. The partial width decaying into the $\alpha + {^{12}{\rm C}}(0_2^+)$ channel is completely suppressed, up to $10^{-2}$ eV, due to the very small value of the penetration factor $P_{L=0}(a)$ in Eqs.~(\ref{eq:gamma}). This is caused by the sufficiently small decay energy of $0.28$ MeV into this channel, in spite of the large overlap between this state and $\alpha + {^{12}{\rm C}}(0_2^+)$ wave function with a certain channel radius, as seen in FIG.~\ref{fig:5}(d). In contrast, for the decay into the $\alpha + {^{12}{\rm C}}(0_1^+)$ channel, large penetration is caused by the large decay energy of $7.9$ MeV. Nevertheless, the small overlap of this state with the $\alpha + {^{12}{\rm C}}(0_1^+)$ wave function suppresses the decay into this channel, and as a result, the relatively small width of $0.14$ MeV is realized, indicating that this state is unusually longlived. The partial decay width into the $\alpha + {^{12}{\rm C}}(0_1^+)$ channel is calculated with a channel radius $a=12.0$ fm. This mechanism also explains the small width of the $(0_6^+)_{\rm OCM}$ state, and hence the observed small width of the $0_6^+$ state at $15.1$ MeV as well, which is discussed in Ref.~\cite{funaki_antisymmet}. In the present calculation, we neglect the decay into the other open channel $\alpha + {^{12}{\rm C}}(2_1^+)$, since the THSR wave function of Eq.~(\ref{eq:thsr}) practically cannot have nonzero spin components with respect to the $\alpha$-$\alpha$ relative motions.

\subsection{Overlap with a single THSR wave function}\label{subsec:overlap}

While the THSR wave function represents the physically clear picture as the $4\alpha$ condensation, the $(0_k^+)_{\rm THSR}$ wave functions $\Psi_k$ with $k=1,\cdots,4$ of Eq.~(\ref{eq:hwwf}) are expressed as the superposed ones of the THSR wave function of Eq.~(\ref{eq:thsr}) with different $B$ values. One might therefore suspect that the superposed wave functions of the $(0_k^+)_{\rm THSR}$ states no longer keep the remarkable character of a single component wave function, retained in Eq.~(\ref{eq:thsr}). In order to clarify this point, we first construct orthonormal wave functions from the THSR wave functions as follows:
\begin{equation}
{\widetilde \Psi}_k(B)= C_{k-1}(B) P_{k-1}\Phi_{4\alpha}(B), \ \ (k=1,\cdots,4), \label{eq:38}
\end{equation}
where $P_{k-1}$ is projection operators defined by,
\begin{eqnarray}
&&P_0=1,\  (k=1),\nonumber \\
&&P_{k-1}=1-\sum_{l=1}^{k-1} | \Psi_l \left\rangle \right\langle \Psi_l |,\ (k=2,3,4), \label{eq:39}
\end{eqnarray}
and $C_{k-1}$ is a normalization constant,
\begin{equation}
C_{k-1}(B)= \left\langle P_{k-1}\Phi_{4\alpha}(B) \Big| P_{k-1}\Phi_{4\alpha}(B) \right\rangle ^{-1/2}.
\end{equation}
The wave function ${\widetilde \Psi}_k(B)$ depends on the parameter $B$ and is orthogonal to the wave functions $\Psi_{k^\prime}$ of the $(0_{k^\prime}^+)_{\rm THSR}$ states with $k^\prime\neq k$. 
We then calculate the following squared overlap amplitudes between the wave functions ${\widetilde \Psi_k}(B)$ and $\Psi_k$,
\begin{equation}
\Theta_k(R_0)=\Big| \left\langle \Psi_k \Big| {\widetilde \Psi}_k(B) \right\rangle \Big|^2,\ \ (k=1,\cdots,4).\label{eq:theta}
\end{equation}
Note that $\Theta_k(R_0)$ is written in terms of the parameter $R_0$, which is related to $B$ like $B=\sqrt{b^2+2R_0^2}$.
\begin{figure}[htbp]
\begin{center}
\includegraphics[scale=0.7]{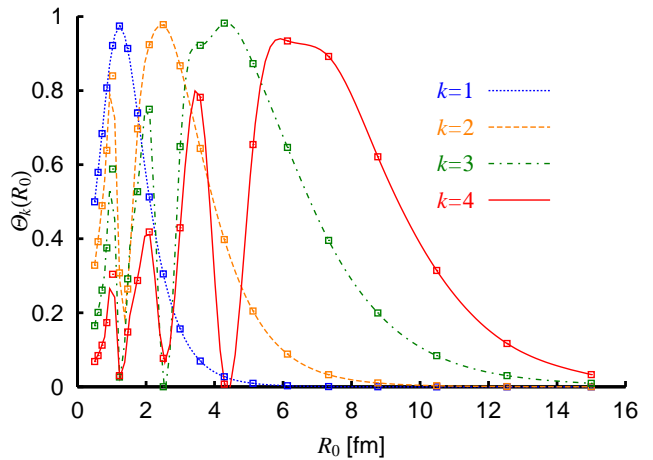}
\caption{(color online). Overlap amplitudes $\Theta_k(R_0)$ between the wave functions $\Psi_k$ and ${\widetilde \Psi}_k(B)$ with $k=1,\cdots,4$, as a function of $R_0$ defined by Eq.~(\ref{eq:theta}).}\label{fig:6}
\end{center}
\end{figure}

In FIG.~\ref{fig:6} we show $\Theta_k(R_0)$ of Eq.~(\ref{eq:theta}) as a function of $R_0$. For any wave functions $\Psi_k$ $(k=1,\cdots,4)$, $\Theta_k(R_0)$ $(k=1,\cdots,4)$ takes values close to 100 \% at optimum values of $R_0$. In particular, the maximum of $\Theta_4(R_0)$ amounts to $96$ \% at $R_0=6.5$ fm. This means that the $(0_4^+)_{\rm THSR}$ state is almost represented by a single THSR wave function parametrized by a large $R_0$ value in the space orthogonal to the other $(0_k^+)_{\rm THSR}$ $(k=1,2,3)$ states. We should note that this feature is exactly verified for the Hoyle state, strongly supporting the picture that the Hoyle state is a $3\alpha$ condensate~\cite{funaki1}. Therefore, the $(0_4^+)_{\rm THSR}$ state shows the $4\alpha$ condensate character, as does the THSR wave function $\Phi_{4\alpha}(R_0)$ with a large $R_0$ value.

The $(0_3^+)_{\rm THSR}$ state also gives a large squared overlap at a large $R_0$ value, i.e. as a maximum $\Theta_3(R_0)=0.98$ at $R_0=4.0$ fm. The r.m.s. radius of $4.2$ fm of this state is comparable to the one of the Hoyle state, and therefore this state also shows $4\alpha$ condensate character. 
The $(0_1^+)_{\rm THSR}$ and $(0_2^+)_{\rm THSR}$ states also have large squared overlaps, with maximal values of $\Theta_1(R_0)=0.98$ and $\Theta_2(R_0)=0.98$ at $R_0=1.2$ fm and $R_0=2.5$ fm, respectively. However, as shown in Table~\ref{tab:1}, the r.m.s. radii of these states are rather small. In these states the antisymmetrizer strongly disturbs the product structure of $4\alpha$ particles, and therefore these rather compact states will not have a dilute gas-like structure of the $4\alpha$'s.

\subsection{Boson mapping of the THSR wave function}\label{subsec:boson}

The quantities such as occupation probability of a single-$\alpha$ orbit and momentum distribution of $\alpha$ particles are very important information to discuss the nature of the $\alpha$ condensate. However, the wave functions $\Psi_k$ of the $(0_k^+)_{\rm THSR}$ states are not based on the degree of freedom of the $\alpha$ particles (boson) but on that of the nucleon (fermion), and therefore the various bosonic quantities cannot directly be given by the wave functions. In order to obtain them, it is necessary to extract the bosonic wave functions from the fermionic wave functions $\Psi_k$.

The wave functions $\Psi_k$ in Eq.~(\ref{eq:hwwf}) are rewritten as follows:
\begin{equation}
\Psi_k={\cal A}\left[\chi_k(\vc{\xi}) \phi_{\alpha_1}\cdots \phi_{\alpha_4} \right], \label{eq:rgmwf}
\end{equation}
and
\begin{eqnarray}
\chi_k(\vc{\xi})\equiv &&\hspace{-4mm}\sum_m f_k(B^{(m)}) \chi_{4\alpha}^{\rm THSR}(B^{(m)};\vc{R}_1,\cdots,\vc{R}_4)\nonumber \\
=&&\hspace{-4mm}\sum_m f_k(B^{(m)})\prod_{i=1}^4 \exp \left\{ -\frac{2}{(B^{(m)})^2} \left( \vc{R}_i-\vc{X}_G \right)^2  \right\} \nonumber \\
=&&\hspace{-4mm}\sum_m f_k(B^{(m)})\prod_{i=1}^3 \exp \left( -\frac{2}{(B^{(m)})^2}\frac{i}{i+1}\vc{\xi}_i^2  \right), \label{eq:chi}
\end{eqnarray}
where $\chi_k$ is defined in terms of Jacobi coordinates: $\vc{\xi}_1=\vc{R}_2-\vc{R}_1$, $\vc{\xi}_2=\vc{R}_3-(\vc{R}_1+\vc{R}_2)/2$, $\vc{\xi}_3=\vc{R}_4-(\vc{R}_1+\vc{R}_2+\vc{R}_3)/3$. We here mention a few features of the wave function Eq.~(\ref{eq:rgmwf}). It can be rewritten in the form
\begin{equation}
\Psi_k=\int d^3\vc{\eta} {\cal A} \left[\delta^3(\vc{\eta}-\vc{\xi}) \phi^4_\alpha \right] \chi_k(\vc{\eta}),
\end{equation}
where the following abbreviations are used: $d^3\vc{\eta}\equiv d\vc{\eta}_1d\vc{\eta}_2d\vc{\eta}_3$, 
$\delta^3(\vc{\eta}-\vc{\xi})\equiv \delta(\vc{\eta}_1-\vc{\xi}_1)\delta(\vc{\eta}_2-\vc{\xi}_2)\delta(\vc{\eta}_3-\vc{\xi}_3)$, and $\phi^4_\alpha\equiv \phi_{\alpha_1}\cdots \phi_{\alpha_4}$. By using the norm kernel~\cite{saito_supl,horiuchi_rgm} defined by
\begin{equation}
{\cal N} (\vc{\xi},\vc{\xi}^\prime)\equiv\frac{(4!)^4}{16!4!}
\left\langle {\cal A} \left[\delta^3(\vc{\eta}-\vc{\xi}) \phi^4_\alpha \right] \Big| {\cal A} \left[\delta^3(\vc{\eta}-\vc{\xi}^\prime) \phi^4_\alpha \right] \right\rangle,
\end{equation}
the orthonormal condition can be expressed as follows:
\begin{equation}
\frac{16! 4!}{(4!)^4}\int d^3\vc{\xi}d^3\vc{\xi}^\prime \chi_k(\vc{\xi}){\cal N}(\vc{\xi},\vc{\xi}^\prime)\chi_{k^\prime} (\vc{\xi}^\prime)
=\left\langle \Psi_k | \Psi_{k^\prime} \right\rangle = \delta_{kk^\prime},
\end{equation}
with $d^3\vc{\xi} (d^3\vc{\xi}^\prime) \equiv d\vc{\xi}_1 d\vc{\xi}_2 d\vc{\xi}_3 (d\vc{\xi}^\prime_1 d\vc{\xi}^\prime_2 d\vc{\xi}^\prime_3)$.
The above relation suggests that, as discussed by several authors~\cite{takahashi,matsumura,yamada_12C}, a normalized wave function in the bosonic space corresponding to the fermionic wave function $\Psi_k$ can be taken as 
\begin{eqnarray}
\psi_k(\vc{\xi})&&=\sqrt{\frac{16! 4!}{(4!)^4}}\int d^3 \vc{\xi}^\prime {\cal N}^{1/2}(\vc{\xi},\vc{\xi}^\prime)\chi_k(\vc{\xi}^\prime) \nonumber \\ 
&&\equiv \sqrt{\frac{16! 4!}{(4!)^4}}{\cal N}^{1/2}\chi_k.\label{eq:21}
\end{eqnarray}
We should note that the above wave function is written in terms of the relative coordinates of the $\alpha$ particles, $\vc{\xi}$, and does not depend on the internal coordinates of $\alpha$ particles, which are integrated out in the norm kernel ${\cal N}(\vc{\xi},\vc{\xi}^\prime)$.

For simplicity, in Refs.~\cite{takahashi,matsumura} the following approximate form has been taken as $\psi_k(\vc{\xi})$, 
\begin{equation}
\psi_k(\vc{\xi})\approx \frac{{\cal N}\chi_k}{\sqrt{\left\langle {\cal N} \chi_k | {\cal N} \chi_k \right\rangle }}. \label{eq:23}
\end{equation}
We also adopt this approximation in the following discussions as one option.

On the other hand, the square root of the norm kernel can be expanded in using the so-called exchange kernel ${\cal K}$~\cite{saito_supl,horiuchi_rgm},
\begin{equation}
{\cal N}^{1/2}=\left( 1-{\cal K} \right)^{1/2}=1-\frac{1}{2}{\cal K}-\frac{1}{8}{\cal K}^2 -\cdots. \label{eq:1-k}
\end{equation}
In dilute density states such as expressed by the THSR wave function, contributions from the exchange kernel ${\cal K}$ can be expected to be small. We therefore approximate the above quantity, up to first order, as follows: First let us consider a projection operator $\Lambda$~\cite{saito_supl,horiuchi_rgm} onto the Pauli allowed space in the $4\alpha$ system. Since the norm kernel ${\cal N}$ is defined in the Pauli allowed space, the following relations are satisfied, ${\cal N} = {\cal N}\Lambda = \Lambda {\cal N}= \Lambda{\cal N}\Lambda$, and hence ${\cal N}^{1/2}=\Lambda {\cal N}^{1/2}\Lambda$. These properties should also be kept after the expansion of Eq.~(\ref{eq:1-k}), which therefore should have the following form in the first order approximation:
\begin{eqnarray}
&&{\cal N}^{1/2}= \Lambda \left( 1-{\cal K} \right)^{1/2} \Lambda \approx \Lambda \left( 1-\frac{1}{2}{\cal K} \right) \Lambda \nonumber \\
&&= \Lambda \left(\frac{1}{2} + \frac{1}{2}{\cal N} \right) \Lambda = \frac{1}{2}\left( \Lambda + {\cal N}\right), \label{eq:norm_approx}
\end{eqnarray}
where for the second equality ${\cal K}=1-{\cal N}$ is used. In diluted density states, to good approximation, the operator $\Lambda$ can only act on any pairs of $\alpha$ particles to exclude the Pauli-forbidden, $0S$, $0D$, and $1S$ relative states. We thus adapt $\Lambda$ to the present calculation as follows,
\begin{equation}
\Lambda \approx 1-\frac{1}{6}\sum_{i< j}^4\sum_{s=0S,0D,1S}| u_{s}(\vc{R}_{ij}) \left\rangle \right\langle u_{s}(\vc{R}_{ij}) |,
\end{equation} 
where $u_s(\vc{R}_{ij})$ is the harmonic oscillator wave function in a $0S$, $0D$, or $1S$ state, denoted by $s$, for the $\alpha$-$\alpha$ relative motion.
Eventually, we can also adopt as another form the following approximate normalized wave function in the bosonic space,
\begin{equation}
\psi_k(\vc{\xi}) \approx \frac{\left( \Lambda +{\cal N}\right) \chi_k}{\sqrt{ \left\langle \left(\Lambda + {\cal N} \right)\chi_k \Big| \left( \Lambda + {\cal N} \right)\chi_k \right\rangle  } }. \label{eq:27}
\end{equation}

We adopt the bosonic wave function $\psi_k$, in the forms of Eq.~(\ref{eq:23}) and Eq.~(\ref{eq:27}), for the present calculations of quantities such as the one-body density matrix in discussing the amount of $\alpha$ condensation. Hereafter, we refer to the ways of approximation for $\psi_k$ of Eq.~(\ref{eq:23}) and Eq.~(\ref{eq:27}) as APR1 and APR2, respectively.

By using the above $\alpha$-particle bosonic wave function, the one-body density matrix for the $\alpha$ particle is defined following Refs.~\cite{takahashi,matsumura,yamada_12C,4aocm,fewbody,dm_yamada}, like
\begin{eqnarray}
\begin{array}{c}
\rho_{\alpha}(\vc{r},\vc{r}^\prime)= \big\langle \psi_k \big| \frac{1}{4}\sum_{j=1}^4 \rho_j  \big| \psi_k \big\rangle, \vspace{0.7em} \\
\rho_j= \big| \delta(\vc{R}_j-\vc{X}_G-\vc{r}^\prime)\big\rangle \big\langle \delta(\vc{R}_j-\vc{X}_G-\vc{r}) \big|. \vspace{0.7em}  \\
\end{array}
\label{eq:37}
\end{eqnarray}
Then, by solving the following eigenvalue problem,
\begin{equation}
\int d\vc{r}^\prime \rho_{\alpha}(\vc{r},\vc{r}^\prime) \varphi(\vc{r}^\prime) = \mu \varphi(\vc{r}), \label{eq:29}
\end{equation}
we can obtain the occupation probability $\mu$ and the corresponding single-$\alpha$ orbit $\varphi(\vc{r})$ as the eigenvalue and eigenfunction, respectively. The occupation probability $\mu$ is labeled with the angular momentum $L$ and the quantum number of a positive integer $n_L$, like $L_{n_L}$: In this work, for a single-$\alpha$ orbit with an angular momentum $L$, we denote the largest occupation probability as $L_1$ $(n_L=1)$, the second largest as $L_2$ $(n_L=2)$, the third largest as $L_3$ $(n_L=3)$, etc. (see also Ref.~\cite{foot_occupation}).\begin{figure}[htbp]
\begin{center}
\includegraphics[scale=0.75]{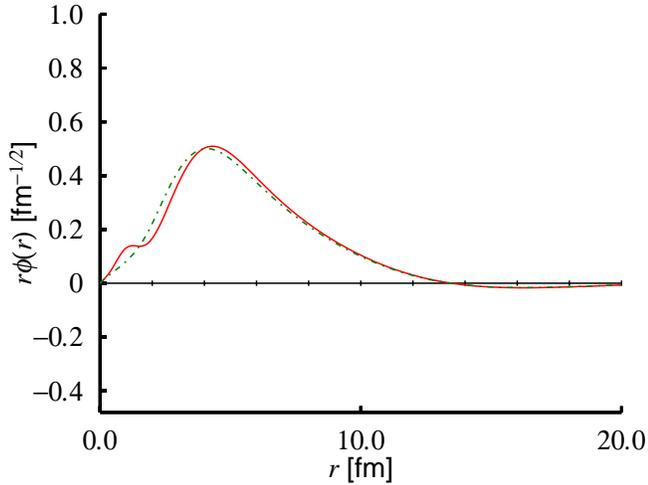}
\caption{(color online). Radial parts of single-$\alpha$-particle orbits with $L=0$ and $n_L=1$ $(S_1)$ for the Hoyle state calculated with APR1 (solid curve) and APR2 (dash-dotted curve) approximations. See text for the definitions of APR1 and APR2.}\label{fig:7}
\end{center}
\end{figure}

\begin{figure}[htbp]
\begin{center}
\includegraphics[scale=0.7]{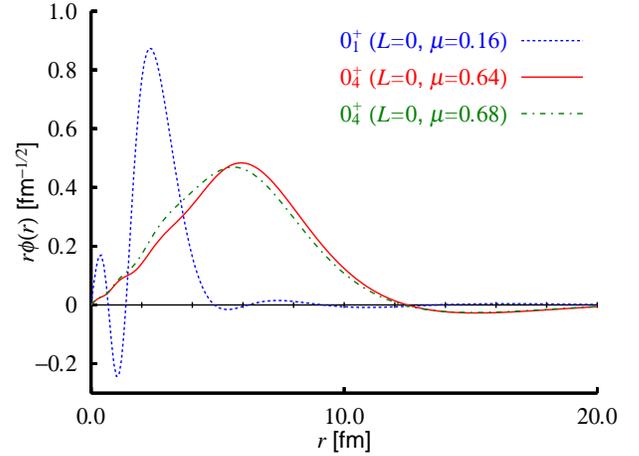}
\caption{(color online). Radial parts of single-$\alpha$-particle orbits with $L=0$ and $n_L=1$ $(S_1)$, for the $(0_1^+)_{\rm THSR}$ (dotted curve) and $(0_4^+)_{\rm THSR}$ (solid and dash-dotted curves) states. See text for the difference between the solid and dash-dotted curves. The values of $\mu$ in the parentheses are the occupation probabilities into the corresponding orbits.}\label{fig:12}
\end{center}
\end{figure}
\begin{figure}[htbp]
\begin{center}
\includegraphics[scale=0.7]{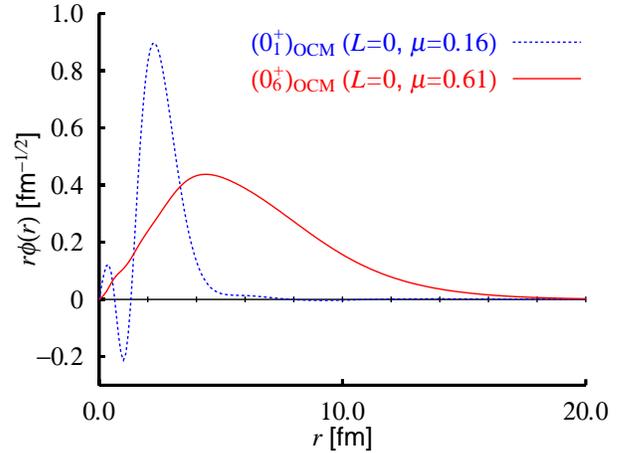}
\caption{(color online). Radial parts of single-$\alpha$-particle orbits with $L=0$ and $n_L=1$ $(S_1)$, for the ground state $(0_1^+)_{\rm OCM}$ (dotted curve) and the Hoyle state $(0_6^+)_{\rm OCM}$ (solid curve) states, which are obtained from the $4\alpha$ OCM calculation~\cite{4aocm}. The values of $\mu$ in the parentheses are the occupation probabilities into the corresponding orbits.}\label{fig:13}
\end{center}
\end{figure}

\begin{figure*}[htbp]
\begin{center}
\includegraphics[scale=0.8]{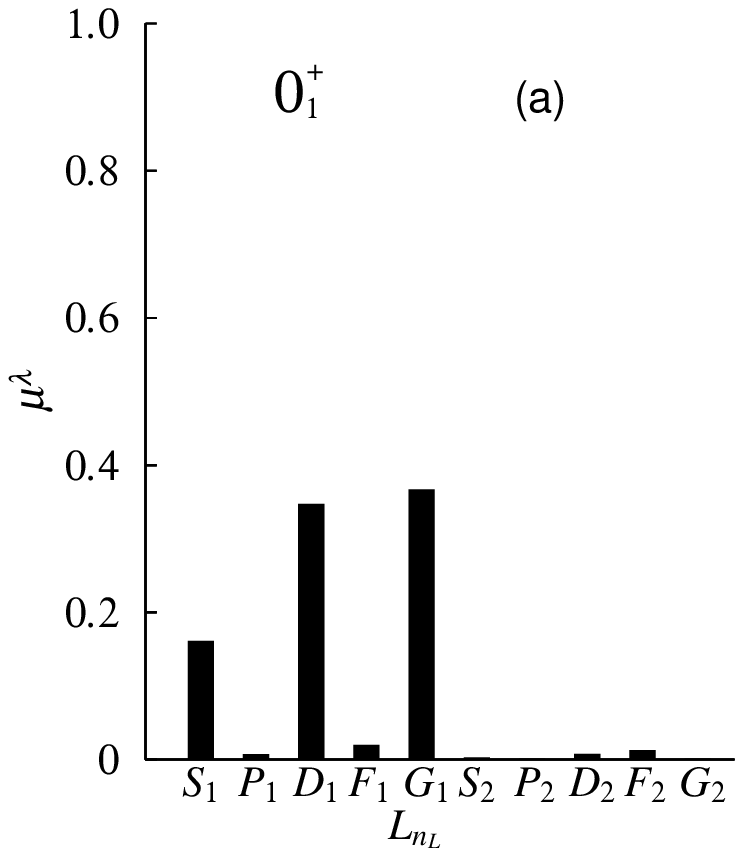}
\includegraphics[scale=0.8]{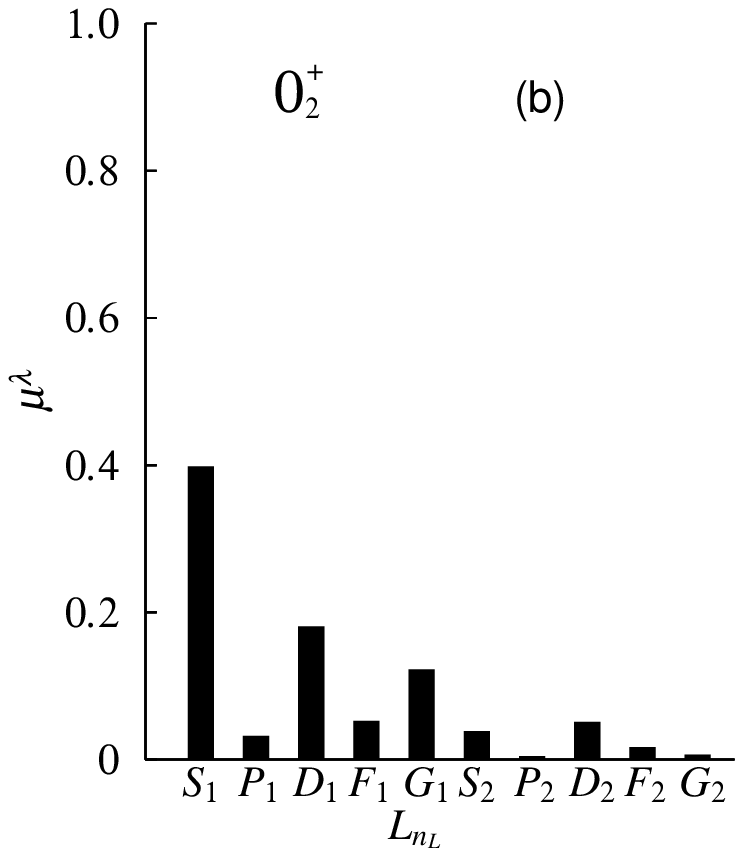}
\end{center}
\begin{center}
\includegraphics[scale=0.8]{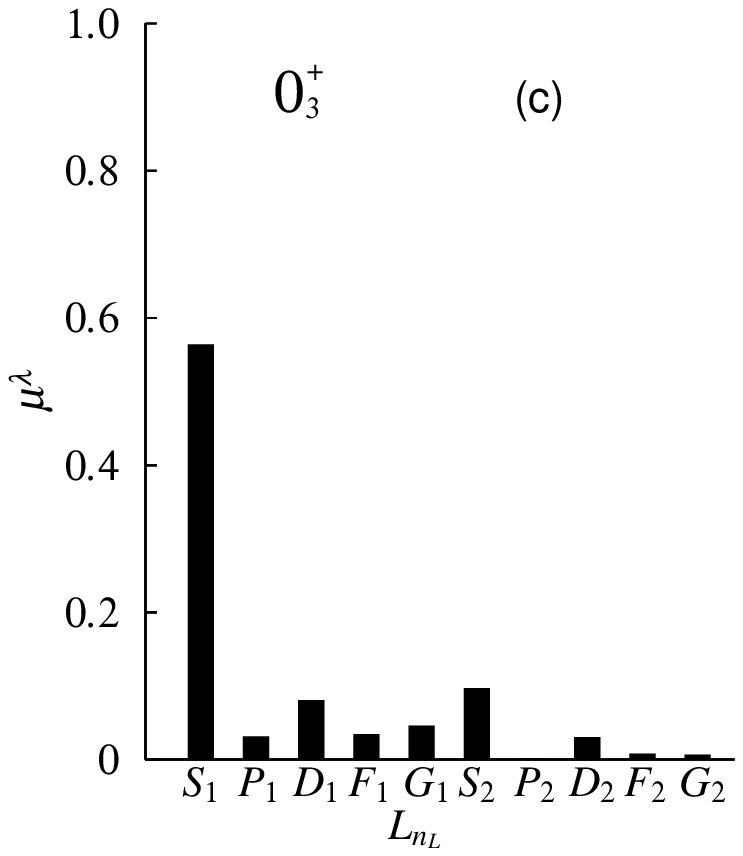}
\includegraphics[scale=0.8]{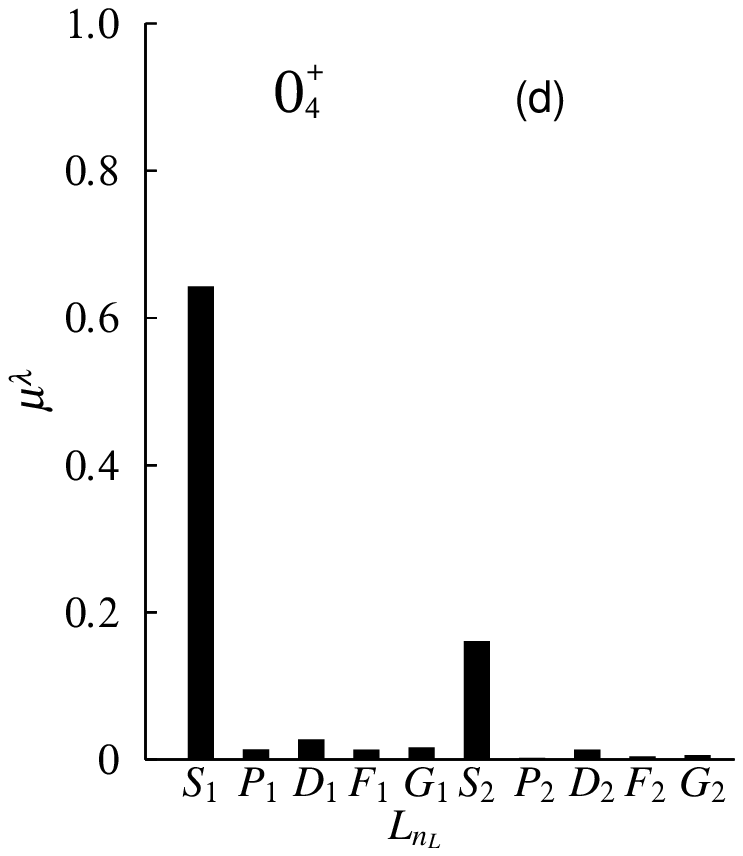}
\end{center}
\caption{Occupation probabilities $\mu$, which are labeled as $L_{n_L}$ with $L$ and $n_L$ being the angular momentum and the quantum number of a positive integer, respectively, for (a) $(0_1^+)_{\rm THSR}$, (b) $(0_2^+)_{\rm THSR}$, (c) $(0_3^+)_{\rm THSR}$, and (d) ($0_4^+)_{\rm THSR}$ states. The approximation of APR1 is adopted to obtain the bosonic wave function.}\label{fig:14}
\end{figure*}

In order to check if the above-mentioned two approximations of APR1 and APR2 work well, at least, in dilute density systems, we first apply the above formalism to the Hoyle state which has already been well known to have the dilute gas-like $3\alpha$ structure. We show in FIG.~\ref{fig:7} the single-$\alpha$ orbits for the Hoyle state, which are labeled as $S_1$ and obtained with the two ways, APR1 (solid curve) and APR2 (dash-dotted curve). The microscopic wave function of the Hoyle state is given by solving the Hill-Wheeler equation with the THSR ansatz, where the same effective nuclear force is adopted as in the present study of ${^{16}{\rm O}}$. We can see that both curves are in good agreement with each other. This fact guarantees that the approximations are realistic for the dilute density state. We see that the single-$\alpha$ $S$ orbit has $0S$ nodal behavior with a very extended tail. This is the characteristic feature of the $\alpha$ condensate state~\cite{yamada_12C,4aocm}, where the $\alpha$ particles occupy the lowest $0S$ orbit of a rather soft mean-field-like potential~\cite{YS}.


As we discussed above for the $^{12}$C case, also for $^{16}$O case it is important to calculate the occupation probability and the corresponding single-$\alpha$-particle orbit, which are the crucial quantities for judging the amount of the $\alpha$ condensation. In FIG.~\ref{fig:12}, the single-$\alpha$ $S$ orbits for the $(0_1^+)_{\rm THSR}$ (dotted curve) and $(0_4^+)_{\rm THSR}$ (solid and dash-dotted curves) states are shown. The solid and dash-dotted curves are obtained with the approximations of APR1 and APR2, respectively, i.e. the ones of Eq.~(\ref{eq:23}) and Eq.~(\ref{eq:27}) for the bosonic wave function of Eq.~(\ref{eq:21}). The dotted curve for the ground state is calculated with APR1. The occupation probabilities into the corresponding orbits are also shown in parentheses.

The fact that both curves with APR1 and APR2 for the $(0_4^+)_{\rm THSR}$ well agree with each other again gives a support that both approximations are taken correctly. The approximation APR2 can be justified for the description of dilute density states, so that the effect of the antisymmetrizer is weakened.

We see in FIG.~\ref{fig:12} for the $(0_4^+)_{\rm THSR}$ state the large single-$\alpha$ $S$ orbit occupancy of $64$ \% and $68$ \% for the approximations APR1 and APR2, respectively, which has a nodeless behavior with a very extended tail. We can see in FIG.~\ref{fig:13} that the $(0_6^+)_{\rm OCM}$ state gives the similar value of $61$ \% of the occupation probability and the corresponding single-$\alpha$ $S$ orbit also resembles the one for the $(0_4^+)_{\rm THSR}$ state, concerning the $0S$ nodal behavior and the spatial extension. These results again give us  support that the $(0_4^+)_{\rm THSR}$ state is to good approximation the counterpart of the $(0_6^+)_{\rm OCM}$ state described in our previous paper \cite{4aocm}. We should note that this behavior of the single-$\alpha$ orbit for this state is characteristic as an analogue to the Hoyle state~\cite{funaki_antisymmet}.

On the other hand, for the ground state, the single-$\alpha$ orbit has $2S$ nodal behavior with a strongly reduced tail, reflecting a compact density and hence the strong effect of the Pauli principle, for both approaches, THSR and OCM.
\begin{figure}[htbp]
\begin{center}
\includegraphics[scale=0.62]{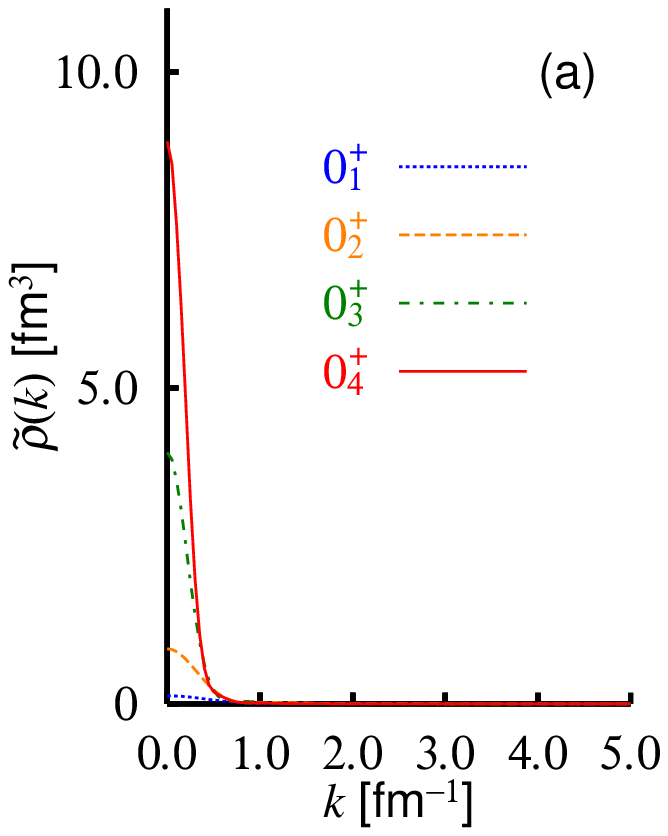}
\includegraphics[scale=0.62]{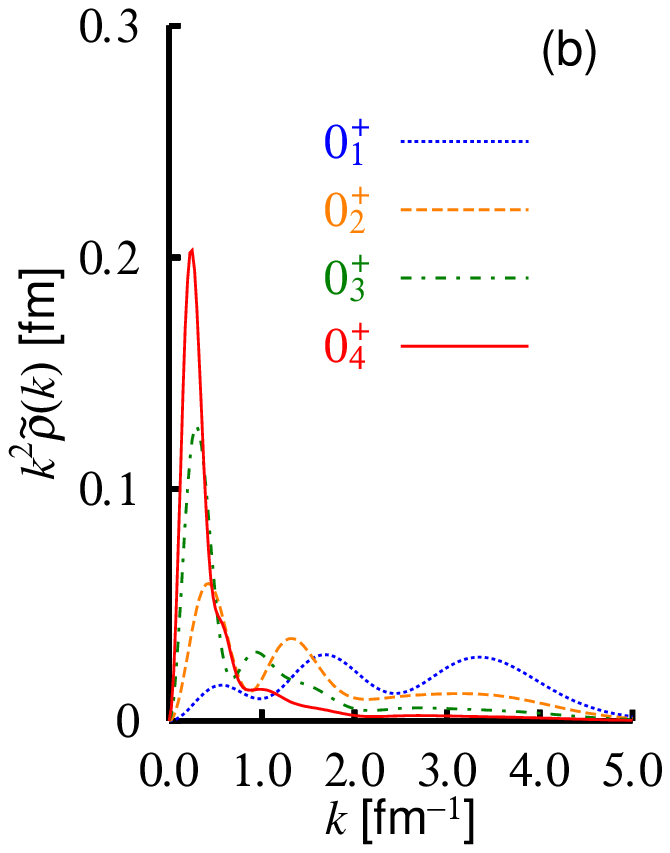}
\caption{(color online). Momentum distributions of the $\alpha$ particle defined by Eq.~(\ref{eq:alpha_mom}), (a) $\rho_\alpha(k)$ and (b) $k^2\rho_\alpha(k)$, for the $(0_1^+)_{\rm THSR}$--$(0_4^+)_{\rm THSR}$ states. The approximation of APR1 is adopted to obtain the bosonic wave function.}\label{fig:15}
\end{center}
\end{figure}
\begin{figure}[htbp]
\begin{center}
\includegraphics[scale=0.615]{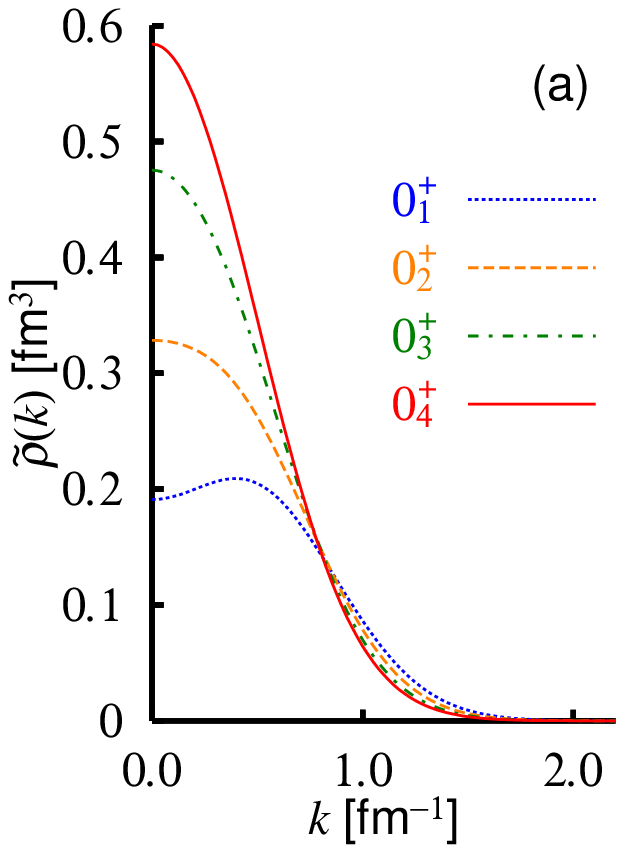}
\includegraphics[scale=0.615]{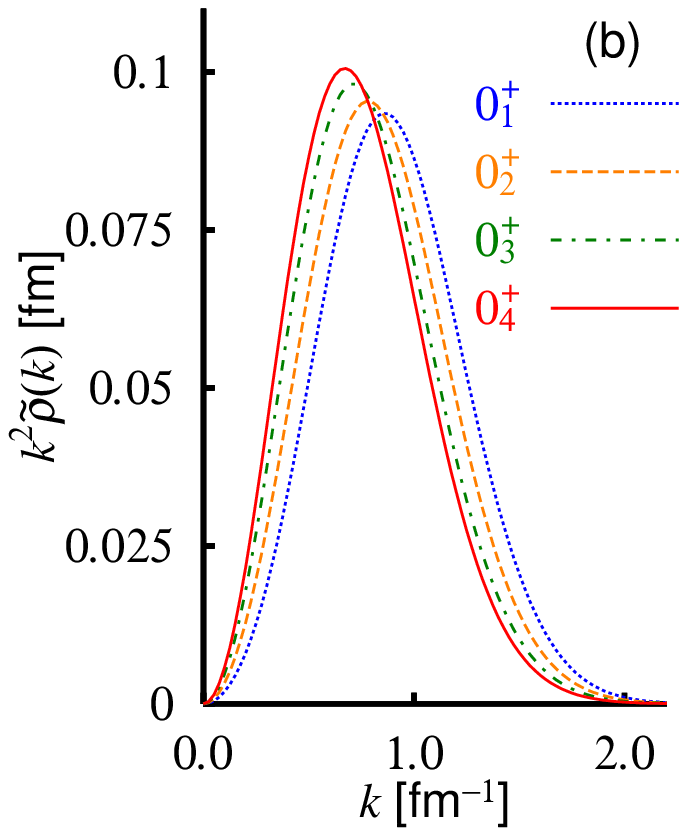}
\caption{(color online). Momentum distributions of the nucleons defined by Eq.~(\ref{eq:density2}), (a) ${\widetilde \varrho}(k)$ and (b) $k^2{\widetilde \varrho}(k)$, for the $(0_1^+)_{\rm THSR}$--$(0_4^+)_{\rm THSR}$ states.}\label{fig:16}
\end{center}
\end{figure}

The distributions of the occupation probabilities into the single $\alpha$ orbits with APR1 are shown for the $(0_1^+)_{\rm THSR}$, $(0_2^+)_{\rm THSR}$, $(0_3^+)_{\rm THSR}$, and $(0_4^+)_{\rm THSR}$ states in FIGS.~\ref{fig:14}(a)--(d), respectively. The orbits shown in FIG.~\ref{fig:12} correspond to the ones labeled as $S_1$ for the $(0_1^+)_{\rm THSR}$ state (FIG.~\ref{fig:14}(a)) and the $(0_4^+)_{\rm THSR}$ state (FIG.~\ref{fig:14}(d)). We can thus see that the occupation probability for the $(0_4^+)_{\rm THSR}$ state concentrates on the single $0S$ orbit shown in FIG.~\ref{fig:12} with $64$ \%. This also means that the remaining $30$ \% to $40$ \% come from the occupations into higher orbits, following the effect of the antisymmetrization. On the other hand, the $(0_1^+)_{\rm THSR}$ state has rather equal distributions of occupation probabilities labeled as $S_1$, $D_1$, and $G_1$. In particular, the largest contribution comes from the $G_1$ orbit, reflecting more or less the doubly closed shell structure with the SU(3) configuration of $[4444](\lambda, \mu)=(0,0)$.

We should mention that the $(0_3^+)_{\rm THSR}$ state and even the $(0_2^+)_{\rm THSR}$ state also have considerable concentration of occupation into a single-$\alpha$ $S$ orbit as seen in FIGS.~\ref{fig:14}(b) and (c). The values of the occupation probability are $40$ \% and $56$ \% for the $(0_2^+)_{\rm THSR}$ and $(0_3^+)_{\rm THSR}$ states, respectively, to be compared with the value of $64$ \% for the $(0_4^+)_{\rm THSR}$ state. However, as mentioned in Sec.~\ref{subsec:wf}, we think that these states, just in a rough average way, correspond to the observed $0_2^+$ and $0_4^+$ states with the $\alpha (S)+ {^{12}{\rm C}}(0_1^+)$ structures, respectively. It is more likely that the concentrations of the occupation into the single-$\alpha$ $S$ orbits for these states disappear in a situation where the $\alpha$-$\alpha$ relative motion is solved in a larger model space so as to include the $\alpha+ ^{12}$C clustering configurations, as was done in the previous $4\alpha$ OCM calculation.

For the $(0_1^+)_{\rm THSR}$--$(0_4^+)_{\rm THSR}$ states, the sums of the occupation probabilities for the ten orbits ($S_1$, $P_1$, $D_1$, $F_1$, $G_1$, $S_2$, $P_2$, $D_2$, $F_2$, and $G_2$) are $0.93$, $0.90$, $0.90$, and $0.90$, respectively. Due to antisymmetrization, there are about 10 percent missed from  higher orbits other than those shown in FIG.~\ref{fig:14}.

The momentum distribution of $\alpha$ particle is also an important quantity to judge which state has a $4\alpha$ condensate nature. It can be obtained by calculating the following doubly Fourier transformation of the one-body density matrix $\rho_\alpha(\vc{r},\vc{r}^\prime)$ in Eq.~(\ref{eq:37}),
\begin{equation}
{\widetilde \rho_{\alpha}}(k)=\frac{1}{(2\pi)^3}\int d\vc{r} d\vc{r}^\prime {\rm e}^{-i\vc{k}\cdot\vc{r}} \rho_{\alpha}(\vc{r},\vc{r}^\prime) {\rm e}^{i\vc{k}\cdot\vc{r}^\prime}. \label{eq:alpha_mom}
\end{equation}
In FIG.~\ref{fig:15} the above momentum distributions of the $\alpha$ particles are shown for the $(0_1^+)_{\rm THSR}$--$(0_4^+)_{\rm THSR}$ states. As is consistent with the large occupation into the single $\alpha$ $0S$ orbit, the $(0_4^+)_{\rm THSR}$ state has the most prominent $\delta$-function-like peak at around zero momentum, which is  one of the typical characters of the $\alpha$ condensate state. According to the rather strong concentrations of the occupation probability on the $S$ orbits for the $(0_2^+)_{\rm THSR}$ and $(0_3^+)_{\rm THSR}$ states, these states also have some enhancement around  zero momentum. On the other hand,  the $(0_1^+)_{\rm THSR}$ state does not show such an enhancement.

We show in FIG.~\ref{fig:16} the nucleon momentum distributions for the $(0_1^+)_{\rm THSR}$--$(0_4^+)_{\rm THSR}$ states, which can be defined, as in the $\alpha$-particle case in Eqs.~(\ref{eq:37}) and (\ref{eq:alpha_mom}), like
\begin{equation}
{\widetilde \varrho}(k)=\frac{1}{(2\pi)^3}\int d\vc{r} d\vc{r}^\prime {\rm e}^{-i\vc{k}\cdot\vc{r}} \varrho(\vc{r},\vc{r}^\prime) {\rm e}^{i\vc{k}\cdot\vc{r}^\prime},
\end{equation}
with
\begin{eqnarray}
\begin{array}{c}
\varrho(\vc{r},\vc{r}^\prime)= \big\langle \Psi_k \big|\frac{1}{16}\sum_{i=1}^{16} \varrho_i \big| \Psi_k \big\rangle, \vspace{0.7em} \\
\varrho_i= \big| \delta(\vc{r}_i-\vc{X}_G-\vc{r}^\prime)\big\rangle \big\langle \delta(\vc{r}_i-\vc{X}_G-\vc{r}) \big|. \vspace{0.7em}  \\
\end{array}
\label{eq:density2}
\end{eqnarray}
The prominent peak of the $\alpha$-particle momentum distribution seen at around zero momentum for the $(0_4^+)_{\rm THSR}$ state disappears in the nucleon momentum distribution. This means that the nucleons, on average, have rather high momentum components even for the $(0_4^+)_{\rm THSR}$ state. The nucleon momentum distribution corresponds to the Fourier transformation of the nucleon density distribution, which is defined by Eq.~(\ref{eq:density3}) (see also FIG.~\ref{fig:4}). Thus, this just reflects the diluteness of the nucleon density and does not give the property of the condensate structure composed of the $\alpha$ particles.


\subsection{Size dependence of the occupation probabilities}\label{subsec:size}

\begin{figure}[htbp]
\begin{center}
\includegraphics[scale=0.7]{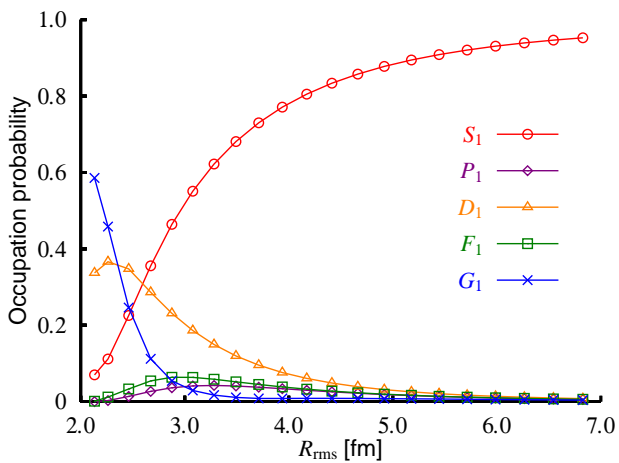}
\caption{(color online). Dependence of the occupation probabilities of the single-$\alpha$-particle orbits labeled as $S_1$, $P_1$, $D_1$, $F_1$, and $G_1$ in the ${^{16}{\rm O}}(0^+)$ state on its r.m.s. radius $R_{\rm rms}$. The way of APR1 is adopted to obtain the bosonic wave function.}\label{fig:18}
\end{center}
\end{figure}

It is instructive to discuss the nuclear size dependence of the occupation probabilities of the single-$\alpha$ orbits in the $^{16}$O$(0^{+})$ state within the framework of the $4\alpha$ THSR ansatz. The results are shown in FIG.~\ref{fig:18} for the single-$\alpha$ orbits labeled as $S_1$, $P_1$, $D_1$ $F_1$, and $G_1$. They are calculated based on the single $4\alpha$ THSR wave function defined by Eq.~(\ref{eq:thsr}), which has the free parameter $B$ triggering the size of the nucleus. Starting with the single $4\alpha$ THSR wave function, we extract the corresponding bosonic wave function with the approximation of APR1, and then calculate the one-body density matrix of the bosonic state, to obtain the occupation probabilities as the eigenvalues of the one-body density matrix. The obtained occupation probabilities then depend on the $B$-parameter, i.e. the r.m.s. radius $R_{\rm rms}$ of the nucleus. 

We can clearly see that with the increase of the r.m.s. radius $R_{\rm rms}$ the occupation probability comes to concentrate on a single $S$ orbit and the ones into the other orbits get rapidly suppressed. A similar behavior is also seen in the previous analysis for the $^{12}$C$(0^{+})$ state based on the $3\alpha$ OCM~\cite{yamada_12C} and in infinite nuclear matter case~\cite{ropke_density}, where the condensate fraction is enhanced as the density decreases. Here we notice that the r.m.s. radii of the $(0_1^+)_{\rm THSR}$ and $(0_4^+)_{\rm THSR}$ states are calculated as $R_{\rm rms}=2.5$ fm and $R_{\rm rms}=6.1$ fm, respectively. The distributions of the occupation probability at $R_{\rm rms}=2.5$ fm are close to the ones of the full solution of the $(0_1^+)_{\rm THSR}$ state using the Hill-Wheeler equation, which are shown in FIG.~\ref{fig:14}. This is reasonable, since the single component THSR wave function, which coincides with the doubly closed shell wave function in the compact limit, as mentioned in Sec.~\ref{subsec:thsr}, describes the $(0_1^+)_{\rm THSR}$ state very well. We should recall that the squared overlap between them is $0.98$, as discussed in Sec.~\ref{subsec:overlap}. However, the occupation probability of the $S_1$ orbit at $R_{\rm rms}=6.1$ fm amounts to about $90$ \%, which is considerably larger than the value of $64$ \% for the $(0_4^+)_{\rm THSR}$ state. The reason of this difference exists in the fact that the $(0_4^+)_{\rm THSR}$ state can no longer be represented to good approximation by the form of the THSR wave function $\Phi_{4\alpha}(B)$ of Eq.~(\ref{eq:thsr}), but by the form of the wave function ${\widetilde \Psi}_{k=4}(R_0)$, as precisely defined by Eq.~(\ref{eq:38}). Let us now recall the discussion made in Sec.~\ref{subsec:overlap}, where the latter wave function is constructed with the orthogonalization to the lower excited states and the ground state. That is, we can think that the orthogonalization reduces the occupation probability into a single $S$ orbit to the value of $64$ \%, from about $90$ \%. In other words, we can say that the reduction is caused by the effect of the antisymmetrization, which remains in the orthogonalization operator $P_{k-1=3}$ defined by Eq.~(\ref{eq:39}).

\subsection{The $(0_2^+)_{\rm THSR}$ and $(0_3^+)_{\rm THSR}$ states in $^{16}$O}\label{subsec:interp}

In the previous sections, we clarified that the $(0_4^+)_{\rm THSR}$ state apparently has a $4\alpha$ condensate structure, and is likely to be the counterpart of the $(0_6^+)_{\rm OCM}$ state with the $4\alpha$ condensate structure obtained in the $4\alpha$ OCM calculation. We therefore consider that the $(0_4^+)_{\rm THSR}$ state probably corresponds to the observed $0_6^+$ state at $15.1$ MeV, as does the $(0_6^+)_{\rm OCM}$ state. In this situation, we should keep in mind that the THSR wave function is not adapted to treat $\alpha+ ^{12}$C configurations, since all $\alpha$'s are treated on equal footing. Therefore, the $(0_2^+)_{\rm THSR}$ and $(0_3^+)_{\rm THSR}$ states obtained from the THSR wave function mock up in an incomplete way the $\alpha+^{12}$C configurations in $^{16}$O. It will be important in future work to include those configurations into the THSR description.

On the other hand, the position of the $(0_3^+)_{\rm THSR}$ state is rather close to the $4\alpha$ threshold, and we see that the state also has a large amount of  $\alpha$ condensate fraction, though the amount is not so large as that of the $(0_4^+)_{\rm THSR}$ state. In this respect, the $(0_3^+)_{\rm THSR}$ state  could also be a candidate for the $4\alpha$ condensate. However, as mentioned above, the result of the $4\alpha$ OCM calculation indicates that a sizable fraction of the $4\alpha$ condensation is only included in the $(0_6^+)_{\rm OCM}$ state and not in the other five $0^+$ states. The explanation may again reside in the fact that we treat with THSR ansatz all $\alpha$ particles equally and then a relatively large r.m.s. radius value necessarily also leads to an $\alpha$-gas-like state whereas in reality a large size also can be obtained from an $\alpha + ^{12}$C configuration. This points to the necessity to include $\alpha + ^{12}$C configurations into the THSR wave function.

Let us mention that in Ref.~\cite{wakasa} the magnitude of the angular distribution of cross section in the $(\alpha,\alpha ^\prime)$ inelastic scattering to the observed $0_4^+$ state agrees very well with that calculated for the $(0_3^+)_{\rm THSR}$ state, in spite of our present conclusion that the $0_3^+$ state is not very similar to the $(0_4^+)_{\rm THSR}$ state. This agreement, nevertheless, seems rather natural, since as discussed in Refs.~\cite{takashina,wakasa}, the magnitude of the angular distribution in the inelastic scattering depends most sensitively on the size of the excited state and not so much on its internal structure.

\section{Conclusion}\label{sec:conc}

In this work, we reanalyzed the THSR wave function for the $4\alpha$ system introduced previously by the last four authors of the present paper.

Searching for states in the continuum we found an additional $0^+$ state. This $(0_4^+)_{\rm THSR}$ state was obtained at 3 MeV above the $4\alpha$ threshold and was shown to have a large r.m.s. radius of 6.1 fm. This finding could be done with the help of the proper treatment of resonances based on the ACCC method, which was developed by some of the present authors. We compared the wave function of the $(0_4^+)_{\rm THSR}$ state with the one of the $(0_6^+)_{\rm OCM}$ state, which was previously obtained with the $4\alpha$ OCM calculation as the $4\alpha$ condensate state. It was clarified that we could reasonably understand that the $(0_4^+)_{\rm THSR}$ state corresponds to the $(0_6^+)_{\rm OCM}$ state, and then is the most appropriate for the $4\alpha$ condensate state, rather than the $(0_3^+)_{\rm THSR}$ state which had been considered to correspond to the condensate state in \cite{thsr,wakasa,funaki_mpla}.

In the analyses of the THSR wave function, we showed that the $(0_4^+)_{\rm THSR}$ state contains a large component of $\alpha + ^{12}$C$(0_2^+)$ clustering in a way which is similar to  the $(0_6^+)_{\rm OCM}$ state. This indicates that both states correspond to each other and have a gas-like configuration of the $4\alpha$ particles as the analogue to the Hoyle state, from the fact that the Hoyle state has a $3\alpha$-gas-like configuration as the $3\alpha$ condensate state. The $(0_4^+)_{\rm THSR}$ state is then described by a single THSR wave function with a large value of the $B$ parameter in a space orthogonal to the other three $(0_1^+)_{\rm THSR}$--$(0_3^+)_{\rm THSR}$ states, though the $(0_1^+)_{\rm THSR}$--$(0_4^+)_{\rm THSR}$ states are in general expressed as the form of superposition of many THSR wave functions with respect to values of the $B$ parameter. In order to further discuss quantitatively the amount of  $\alpha$ condensate fraction, we extracted the boson degrees of freedom from the THSR wave function, which is microscopically described based on the nucleon degrees of freedom. It was done via somewhat approximate procedures which should be valid at low densities. We showed that the occupation probability for the $(0_4^+)_{\rm THSR}$ state concentrates by a large amount on a single-$\alpha$ $0S$ orbit, like the $(0_6^+)_{\rm OCM}$ state. The shape of the orbits for both states is similar to each other and characteristic for Hoyle-like states. Furthermore, the $(0_4^+)_{\rm THSR}$ state is shown to have a strong peak around zero momentum in the $\alpha$-particle momentum distribution, indicating that the $\alpha$ particles are condensed around zero momentum. We also discussed the reason why the $(0_3^+)_{\rm THSR}$ state, as well as the $(0_2^+)_{\rm THSR}$ state, cannot be regarded as corresponding to the $4\alpha$ condensate state, in spite of the fact that the states possess somewhat enhanced condensate-like features.

All these results strongly suggest that the $(0_4^+)_{\rm THSR}$ state corresponds to the $(0_6^+)_{\rm OCM}$ state and is the most appropriately considered to be the $4\alpha$ condensate state. This further gives us a strong support that the $4\alpha$ condensate state exists around the $4\alpha$ breakup threshold and is very likely the observed $0_6^+$ state at $15.1$ MeV.

This conclusion will be supplemented with more extended calculation than the present THSR description so as to incorporate $\alpha + {^{12}{\rm C}}$ clustering configurations. As concluded from comparison with OCM calculations, the $\alpha + {^{12}{\rm C}}$ components may then have a substantial contribution so that the condensate fractions will be reduced. It is nevertheless expected that the $(0_4^+)_{\rm THSR}$ state remains an excellent approximation also in such an extended THSR description.

\acknowledgments

One of the authors (Y. F.) acknowledges financial assistance from the Special Postdoctoral Researchers Program of RIKEN. This work was partially supported by JSPS (Japan Society for the Promotion of Science) Grant-in-Aid for Young Scientists (B) (21740209) and for Scientific Research (C) (21540283).


\begin{thebibliography}{99}
\bibitem{tang}
K. Wildermuth and Y. C. Tang, {\it A Unified Theory of the Nucleus} (Vieweg, Braunschweig, 1977).

\bibitem{ikeda68}
K. Ikeda, H. Horiuchi, and S. Saito, Prog. Theor. Phys. Suppl. {\bf 68}, 1 (1980).

\bibitem{shell}
A. Arima, H. Horiuchi and T. Sebe, Phys. Lett. {\bf 24 B}, 129 (1967).


\bibitem{Suz76}
Y. Suzuki, Prog. Theor. Phys. {\bf 55}, 1751 (1976); {\bf 56}, 111 (1976).

\bibitem{baye2}
M. Libert-Heinemann, D. Baye, P. -H. Heenen, Nucl. Phys. A {\bf 339}, 429 (1980).

\bibitem{Kat92}
K.~Fukatsu and K. Kat${\rm{\bar{o}}}$, Prog. Theor. Phys. {\bf 87}, 151 (1992).

\bibitem{Ike68Hor72}
K. Ikeda, N. Takigawa and H. Horiuchi, Prog. Theor. Phys. Suppl. Extra Number (1968), 464; H. Horiuchi, K. Ikeda and Y. Suzuki, Prog. Theor. Phys. Suppl. {\bf 52}, 89 (1972).

\bibitem{morinaga}
H. Morinaga, Phys. Rev. {\bf 101}, 254 (1956); Phys. Lett. {\bf 21} 78 (1966).

\bibitem{chevallier}
P. Chevallier, F. Scheibling, G. Goldring, I. Plesser, and M. W. Sachs, Phys. Rev. {\bf 160}, 827 (1967).

\bibitem{freer_chain}
M. Freer {\it et al.}, Phys. Rev. C {\bf 51}, 1682 (1995); {\bf 70}, 064311 (2004).

\bibitem{thsr}
A.~Tohsaki, H. Horiuchi, P. Schuck and G. R\"opke, Phys. Rev. Lett. {\bf 87}, 192501 (2001).

\bibitem{carbon}
For example, Y. Fujiwara, H. Horiuchi, K. Ikeda, M. Kamimura, K. Kat${\rm{\bar{o}}}$, Y. Suzuki, and E. Uegaki, Prog. Theor. Phys. Suppl. {\bf 68}, 29 (1980).

\bibitem{funaki1}
Y. Funaki, A. Tohsaki, H. Horiuchi, P. Schuck and G. R\"opke, Phys. Rev. C {\bf 67}, 051306(R) (2003).

\bibitem{matsumura}
H. Matsumura and Y. Suzuki, Nucl. Phys. {\bf A 739}, 238 (2004).

\bibitem{yamada_12C}
T. Yamada and P. Schuck, Eur. Phys. J. A, {\bf 26}, 185 (2005).

\bibitem{YS}
T. Yamada and P. Schuck, Phys. Rev. C {\bf 69}, 024309 (2004).

\bibitem{kamimura}
Y. Fukushima and M. Kamimura, {\it{Proc. Int. Conf. on Nuclear Structure}}, Tokyo, 1977, ed. T. Marumori (Suppl. of J. Phys. Soc. Japan, {\bf 44}, 225 (1978)); M. Kamimura, Nucl. Phys. A {\bf 351}, 456 (1981).

\bibitem{funaki_8Be}
Y. Funaki, H. Horiuchi, A. Tohsaki, P. Schuck, and G. R\"opke, Prog. Theor. Phys. {\bf 108}, 297 (2002).

\bibitem{roepke}
G. R\"opke, A. Schnell, P. Schuck, and P. Nozieres, Phys. Rev. Lett. {\bf 80}, 3177 (1998).

\bibitem{beyer}
M. Beyer, S. A. Sofianos, C. Kuhrts, G. R\"opke, and P. Schuck, Phys. Lett. B {\bf 488}, 247 (2000). 

\bibitem{slr09}
T. Sogo, R. Lazauskas, G. R\"opke, and P. Schuck, Phys. Rev. C {\bf 79}, 051301 (2009).

\bibitem{itoh}
M. Itoh {\it et al.}, Nucl. Phys. A {\bf 738}, 268 (2004).

\bibitem{koka}
Tz. Kokalova {\it et al.}, Eur. Phys. J A {\bf 23}, 19 (2005); Phys. Rev. Lett. {\bf 96}, 192502 (2006).

\bibitem{freer}
M. Freer {\it et al.}, Phys. Rev. C {\bf 71}, 047305 (2005); {\bf 76}, 034320 (2007); {\bf 80}, 041303(R) (2009).

\bibitem{ohkubo}
S. Ohkubo and Y. Hirabayashi, Phys. Rev. C {\bf 70}, 041602(R) (2004); {\bf 75}, 044609 (2007); Phys. Lett. B {\bf 684}, 127 (2010).

\bibitem{funaki_res}
Y. Funaki, A. Tohsaki, H. Horiuchi, P. Schuck and G. R\"opke, Eur. Phys. J. A {\bf 24}, 321 (2005); {\bf 28}, 259 (2006).

\bibitem{takashina}
M. Takashina and Y. Sakuragi, Phys. Rev. C {\bf 74}, 054606 (2006); M. Takashina, Phys. Rev. C {\bf 78}, 014602 (2008).

\bibitem{kurokawa}
C. Kurokawa and K. Kat${\rm{\bar{o}}}$, Phys. Rev. C {\bf 71}, 021301 (2005); Nucl. Phys. A {\bf 792}, 87 (2007).

\bibitem{neff}
M. Chernykh, H. Feldmeier, T. Neff, P. von Neumann-Cosel, A. Richter, Phys. Rev. Lett. {\bf 98}, 032501 (2007).

\bibitem{enyo_12C}
Y. Kanada-En'yo, Prog. Theor. Phys. {\bf 117}, 655 (2007). 

\bibitem{kawabata}
T. Kawabata {\it et al.}, Phys. Lett. B {\bf 646}, 6 (2007).

\bibitem{enyo_11B}
Y. Kanada-En'yo, Phys. Rev. C {\bf 75}, 024302 (2007).

\bibitem{yamada_11B}
T. Yamada and Y. Funaki, Int. J. Mod. Phys. E {\bf 17}, 2101 (2008).

\bibitem{wakasa}
T.~Wakasa {\it et al.}, Phys Lett B {\bf 653}, 173 (2007).

\bibitem{funaki_mpla}
Y. Funaki, A. Tohsaki, H. Horiuchi, P. Schuck and G. R\"opke, Mod. Phys. Lett. A {\bf 21}, 2331 (2006).

\bibitem{4aocm}
Y. Funaki, T. Yamada, H. Hoiuchi, G. R\"opke, P. Schuck and A. Tohsaki, Phys. Rev. Lett. {\bf 101}, 081502 (2008).

\bibitem{GEM}
M. Kamimura, Phys. Rev. A {\bf 38}, 621 (1988); E. Hiyama Y. Kino, and M. Kamimura, Prog. Part. Nucl. Phys. {\bf 51}, 223 (2003).

\bibitem{monopole}
T. Yamada, Y. Funaki, H. Horiuchi, K. Ikeda, and A. Tohsaki, Prog. Theor. Phys. {\bf 120}, 1139 (2008).

\bibitem{f1}
A. Tohsaki, Phys. Rev. C {\bf 49}, 1814 (1994).

\bibitem{rmsmin}
Y. Funaki, H. Horiuchi, and A. Tohsaki, Prog. Theor. Phys. {\bf 115}, 115 (2006).

\bibitem{accc} V. I. Kukulin and V. M. Krasnopol'sky, J. of Phys. A
{\bf 10}, 33 (1977); V. I. Kukulin, V. M. Krasnopol'sky and M. Miselkhi, Sov. J. Nucl. Phys. {\bf 29}, 421 (1979); V. I. Kukulin, V. M. Krasnopol'sky, and J. Hor$\acute{\rm a}\check{\rm c}$ek, {\it The Theory of Resonances: Principles and Applications} (Kluwer Academic, Dortrecht, 1983), Chapter 5.

\bibitem{TSVL} N. Tanaka, Y. Suzuki and K. Varga, Phys. Rev. C {\bf 54}, 562 (1997); N. Tanaka, Y. Suzuki, K. Varga and R. Lovas, Phys. Rev. C {\bf 59}, 1391 (1999).

\bibitem{Ring/Schuck}
P. Ring and P. Schuck, {\it The Nuclear Many Body Problem} (Springer-Verlag, New York, 1980).

\bibitem{ajze}
F. Ajzenberg-Selove, Nucl. Phys. A {\bf 46}, 1 (1986).

\bibitem{funaki_antisymmet}
Y. Funaki, H. Horiuchi, W. von Oertzen, G. R\"opke, P. Schuck, A. Tohsaki, and T. Yamada, Phys. Rev. C {\bf 80}, 064326 (2009).

\bibitem{lane}
A. M. Lane and R. G. Thomas, Rev. Mod. Phys. {\bf 30}, 257 (1958).

\bibitem{saito_supl}
S. Saito, Prog. Theor. Phys. Suppl. {\bf 62}, 11 (1977).

\bibitem{horiuchi_rgm}
H. Horiuchi, Prog. Theor. Phys. Suppl. {\bf 62}, 90 (1977).

\bibitem{takahashi}
Y. Suzuki and M. Takahashi, Phys. Rev. C {\bf 65}, 064318, (2002).

\bibitem{fewbody}
Y. Suzuki, W. Horiuchi, M. Orabi, K. Arai, Few-Body Systems {\bf 42}, 33 (2008).
\bibitem{dm_yamada}
T. Yamada, Y. Funaki, H. Horiuchi, G. R\"opke, P. Schuck and A. Tohsaki, Phys. Rev. A {\bf 78}, 035603 (2008); Phys. Rev. C {\bf 79}, 054314 (2009).

\bibitem{foot_occupation}
Note that the positive integer $n_L$ is different from the number of node for the radial part of the corresponding single-$\alpha$ orbit $\varphi(\vc{r})$. For instance, in FIG.~\ref{fig:12}, the single-$\alpha$ orbits labeled as $L=0$ and $n_L=1$ $(S_1)$ have $2S$ and $0S$ nodal behaviour for the ground $(0_1^+)_{\rm THSR}$ and the $(0_4^+)_{\rm THSR}$ states, respectively.

\bibitem{ropke_density}
Y. Funaki, H. Horiuchi, G. R\"opke, P. Schuck, A. Tohsaki, and T. Yamada, Phys. Rev. C {\bf 77}, 064312 (2008).

\end{thebibliography}
\end{document}